\documentclass[fleqn,usenatbib]{mnras}

\usepackage{newtxtext,newtxmath}
\usepackage[T1]{fontenc}
\usepackage{ae,aecompl}
\usepackage[dvipsnames]{xcolor}

\usepackage{graphicx}	% Including figure files

\makeatletter
\DeclareRobustCommand{\ion}[2]{%
  \text{#1\,\check@mathfonts\fontsize\sf@size\z@\selectfont #2}%
}
\makeatother

\newcommand{\hi}{\ion{H}{I}}

\newcommand{\ci}{\ion{C}{I}}

\title[Determining the PDR properties of the ISM using {\sc PDFchem}]{{\sc PDFchem}: A new fast method to determine ISM properties and infer environmental parameters using probability distributions}%A new algorithm to determine the photodissociation region properties of the ISM using probability distributions}

\author[Thomas G. Bisbas et al.]{Thomas G. Bisbas$^{1,2,3}$\thanks{E-mail: tbisbas@zhejianglab.com (TGB)},
Ewine F. van Dishoeck$^{4,5}$,
Chia-Yu Hu$^{6,4}$,
and Andreas Schruba$^{4}$
\\
$^{1}$Research Center for Intelligence Computing Platforms, Zhejiang Laboratory, Hangzhou 311100, China\\
$^{2}$I. Physikalisches Institut, Universit\"at zu K\"oln, Z\"ulpicher Stra\ss e 77, K\"oln, Germany\\
$^{3}$Department of Physics, Aristotle University of Thessaloniki, GR-54124 Thessaloniki, Greece\\
$^{4}$Max-Planck-Institut f\"ur Extraterrestrische Physik, Giessenbachstrasse 1, D-85748 Garching, Germany\\
$^{5}$Leiden Observatory, Leiden University, P.O. Box 9513, NL-2300 RA Leiden, the Netherlands\\
$^{6}$Department of Astronomy, University of Florida, Gainesville, FL 32611, USA
}

\date{Accepted XXX. Received YYY; in original form ZZZ}

\pubyear{2021}

\begin{document}
\label{firstpage}
\pagerange{\pageref{firstpage}--\pageref{lastpage}}
\maketitle

\begin{abstract} 
Determining the atomic and molecular content of the interstellar medium (ISM) as a function of environmental parameters is of fundamental importance to understand the star-formation process across the epochs. Although there exist various three-dimensional hydro-chemical codes modelling the ISM at different scales and redshifts, they are computationally expensive and inefficient for studies over a large parameter space.
%Advanced but computationally expensive three-dimensional hydro-chemical approaches are able to model the ISM at different scales and redshifts providing an insight of this process. On the other hand, the wealth of observations in the ALMA era is disproportionately increasing, creating a gap between the ability of the three-dimensional simulations to model the vast number of emission lines and chemical abundances, and the plethora of spectra taken from various local and extragalactic systems. 
Building on our earlier approach, we present {\sc PDFchem}, a novel algorithm that models the cold ISM at moderate and large scales using functions connecting the quantities of the local (effective, 3D) visual extinction, $A_{\rm V,eff}$, the observed (2D) visual extinction, $A_{\rm V,obs}$ and the local number density, $n_{\rm H}$, with probability density functions (PDF) of $A_{\rm V,obs}$ on cloud scales typically tens-to-hundreds of pc as an input. For any given $A_{\rm V,obs}$-PDF, sampled with thousands of clouds, the algorithm instantly computes the average abundances of the most important species (H{\sc i}, H$_2$, C{\sc ii}, C{\sc i}, CO, OH, OH$^+$, H$_2$O$^+$, CH, HCO$^+$) and performs radiative transfer calculations to estimate the average emission of the most commonly observed lines ([C{\sc ii}]~$158\mu$m, both [C{\sc i}] fine-structure lines and the first five rotational transitions of $^{12}$CO). We examine two $A_{\rm V,obs}$-PDFs corresponding to a non star-forming and a star-forming ISM region, under a variety of environmental parameters combinations. These cover FUV intensities in the range of $\chi/\chi_0=10^{-1}-10^3$, cosmic-ray ionization rates in the range of $\zeta_{\rm CR}=10^{-17}-10^{-13}\,{\rm s}^{-1}$ and metallicities in the range of $Z=0.1-2\,{\rm Z}_{\odot}$. {\sc PDFchem} is fast, easy to use, reproduces the PDR quantities of the time-consuming hydrodynamical models and can be used directly with observed data to understand the evolution of the cold ISM chemistry.
\end{abstract}

\begin{keywords}
\textit{(ISM:)} photodissociation region (PDR) --
radiative transfer --
methods: numerical
\end{keywords}

\defcitealias{Bisb19}{Paper~I}
\defcitealias{Bisb21}{B21}

%%%%%%%%%%%%%%%%%%%%%%%%%%%%%%%%%%%%%%%%%%%%%%%%%%

%%%%%%%%%%%%%%%%% BODY OF PAPER %%%%%%%%%%%%%%%%%%

\section{Introduction}

Star-formation occurs predominantly in the cold ($T_{\rm gas} \lesssim 30\,{\rm K}$) and dense ($n_{\rm H} \gtrsim 10^4\,{\rm cm}^{-3}$) H$_2$-rich gas of the interstellar medium (ISM). To understand its evolution throughout cosmic time, knowledge of the molecular mass content of the ISM in galaxies is needed \citep[see][for a review]{Tacc20}. One of the key issues towards this is to determine the distribution of species in large-scale objects, spanning from many parsec to galactic scales, since this can reveal the state of the ISM as well as its potential fraction that leads to star-formation. In addition, there is evidence that the so-called `star formation rate' (SFR), depends on the environmental parameters controlling the distribution of species and the overall thermal balance of the ISM gas. The SFR is a fundamental parameter describing the cyclic process of global star formation and it is observed to vary across cosmic time \citep[see][for a review]{Mada14}. It peaks at a redshift of $z \sim 2.5$ marking the galaxy assembly epoch, when the H$_2$ mass density also peaks \citep{Genz10,Garr21}. It is also known to be different within our Galaxy when comparing, for instance, the Galactic Centre with the outermost parts \citep[see][for a review]{Kenn12}. The higher the SFR, the more energetic the ISM is considered to be. For example, the unattenuated intensity of the FUV radiation field, $\chi$, is frequently assumed to increase with SFR \citep{Bial20} and the cosmic-ray ionization rate, $\zeta_{\rm CR}$, is also suggested to increase in a similar fashion \citep[e.g.][]{Papa10,Gach19}. 

Advanced three-dimensional \mbox{(magneto-)}\linebreak[0]{}hydrodynamical codes coupled with chemical networks are becoming a powerful way to simulate the dynamics and chemistry of the ISM mimicking the environmental conditions observed in low- and high-redshift galaxies. The inclusion of ISM chemistry in such codes allows to study the atomic-to-molecular transition as well as the different phases of the carbon cycle, defining the so-called photodissociation regions \citep[`PDRs';][]{Tiel85,Holl99,Roll07,Wolf22}. It also allows the construction of synthetic observations \citep[see][for a review]{Hawo18} providing the means to directly compare the simulated clouds with real data. 

A frequently adopted time-dependent chemical network that allows on-the-fly calculations of abundances of the most important species was implemented by \citet{Glov10} who significantly improved earlier attempts of \citet{Nels97,Nels99}. Following a similar approach, the {\sc krome} package \citep{Grass14} was developed offering the possibility to study the chemical evolution on a cosmological scale including early Universe. A growing number of projects are focusing on simulating the evolution of the multiphase ISM at kpc scales using time-dependent chemistry. The `SILCC' project \citep{Walc15,Giri16,Seif17,Seif20} models small but vertically elongated regions of a galactic disk focusing on studies related to the cycling process of star-formation under different environmental conditions. Hydro-chemical simulations of dwarf galaxies have been performed by \citet{Hu16,Hu17} to study the formation of molecular gas in the low-metallicity ISM. \citet{Rich16} modelled isolated galaxies of $10^9\,{\rm M}_{\odot}$ in total mass at a resolution of $750\,{\rm M}_{\odot}$ per particle and examined the emission of C{\sc ii} and CO in non-equilibrium (time-dependent) chemistry and with varying metallicity and UV radiation intensities. The `Cloud Factory' project \citep{Smit20} performs high-resolution galaxy-scale simulations and models the ISM in a typical spiral galaxy at a mass resolution of $0.25\,{\rm M}_{\odot}$ resolving the dynamics and fragmentation of such scales. Recently, \citet{Hu21} examined the effect on the ISM of varying the metallicity in a kpc galactic region at 0.2 pc resolution with vertical extension such as in the SILCC project, most notably how the atomic-to-molecular transition as well as the carbon-cycle abundances are affected.

To reduce the amount of computations needed when solving the chemistry on-the-fly, alternative approaches have been attempted. These include a large grid of pre-calculated astrochemical simulations which are then tabulated and used during each hydrodynamical timestep. Although this approach may limit the ability to study properly the chemistry in turbulent regions where the gas mixing timescale is comparable to the chemical time needed to reach chemical equilibrium \citep{Xie95}, it may well be used to estimate the average abundances and diagnostics in the large-scale ISM. Such a methodology has been adopted in the works of \citet{Wu17} and \citet{Li18} using {\sc pypdr} and {\sc cloudy} \citep{Ferl17} pre-calculated grids to study cloud-cloud collisions and whole galactic-disk ISM, respectively. Recently, \citet{Ploe20} performed a great number of PDR calculations (also using {\sc cloudy}) which have been tabulated in publicly available datacubes for usage in hydrodynamical simulations for ISM studies from local to high-redshifts. Such a methodology has been also used in steady-state calculations of large and high-resolution three-dimensional clouds on pc-scales \citep[e.g.][]{Bisb17b}.

Other approaches for estimating the abundances of species as well as the emission of atomic and molecular lines at a post-processed stage have been also considered. In this way, \citet{Offn13} modelled the distribution of abundances of turbulent star-forming clouds using {\sc 3d-pdr} \citep{Bisb12}. In a follow-up work and combining these results with {\sc radmc-3d} \citep{Dull12}, \citet{Offn14} studied the ability of the [\ci] $^{3}P_{1}{\rightarrow}{^{3}}P_0$ fine-structure line to trace H$_2$ gas as an alternative to $^{12}$CO $J=1{-}0$. Similarly -- albeit on a much larger scale -- \citet{Gong18,Gong20} examined the behavior of the CO-to-H$_2$ conversion factor under different ISM environmental conditions, by post-processing TIGRESS simulations \citep{Kim17} and using the \citet{Gong17} simplified network for calculating the hydrogen and carbon chemistry. Recently, \citet[][hereafter \citetalias{Bisb21}]{Bisb21} post-processed two distributions of $14^3\,{\rm pc}^3$ in volume, which are sub-structures taken from the \citet{Wu17} MHD simulations. Using {\sc 3d-pdr} they studied the PDR diagnostics covering both the distribution of chemical abundances as well as the line emission of the most important coolants under ten different ISM environmental conditions. These conditions were covering FUV intensities in the range of $\chi/\chi_0=1-10^3$, cosmic-ray ionization rates in the range of $\zeta_{\rm CR}=10^{-17}-10^{-14}\,{\rm s}^{-1}$ and metallicities in the range of $Z=0.1-2\,{\rm Z}_{\odot}$.

On the other hand, the observational archive is constantly enriched with ionic, atomic and multi-$J$ transitions of molecular lines, from local clouds to galaxies in the distant Universe \citep[e.g.][]{Beut14,Mash15,Herr18,Bigi20,Henr22}. If we are to compare the ability and efficiency of the aforementioned numerical approaches to model the multiphase ISM under various environmental conditions, with the wealth of observational datacubes taken from {\it Herschel}, SOFIA, ALMA, JVLA, and other space and ground-based telescopes, we will observe a disproportional enrichment of the latter. This is because the high computational demand of simulations does not always allow the study of a large range of various three-dimensional density distributions embedded in different ISM conditions to model a particular object. As a natural consequence, limitations are adopted in which i) the complexity of the density distribution is reduced to one-dimensional uniform-density slab representatives, ii) the chemistry is treated with minimal heating and cooling processes, and iii) the radiative transfer calculations are simplified, lacking of chemistry. Such limitations may lead to degenerate solutions in determining the ISM environmental conditions from observations, all the more since the non-linear nature of chemistry combined with the great number of free parameters defining such an environment (i.e. intensity of FUV radiation, cosmic-ray ionization rate, metallicity, shock heating, X-ray intensity etc.) frames a higher-dimensional problem requiring an in-depth exploration.

The structure of the ISM is observed to be very complex and far from uniform. Revealing its three-dimensional structure is a difficult-to-impossible task. Instead, it is often represented by a characteristic distribution of visual extinctions, the so-called $A_{\rm V}$-PDF. The quantity $A_{\rm V}$ represents the visual extinction which correlates with the total H-nucleus column density, $N_{\rm H}$,  with a conversion factor that depends on on the assumed grain size distribution \citep[e.g.][]{Wein01,Rach09}. In this work, the relation $A_{\rm V} = N_{\rm H}\cdot6.3\times10^{-22}\,{\rm mag}$ is used \citep{Roll07}. Such distributions of $A_{\rm V}$ have been studied in local systems \citep[e.g.][]{Good09,Kain09,Froe10,Abre15,Schn16,Ma22}, but are difficult to determine in the extragalactic context due to resolution limitations \citep[see, however][for such attempts]{Lero16,Sawa18}. They provide an opportunity to use them to mimic the complexity of the ISM structure and calculate their atomic and molecular mass content with much simpler PDR models.

All the above have motivated our earlier efforts discussed in \citet[][hereafter \citetalias{Bisb19}]{Bisb19} in which a new, fast method for calculating the atomic and molecular content of cold ISM regions has been presented. Contrary to the standard way of using \mbox{(non-)}\linebreak[0]{}uniform density distributions as inputs, the new method considers user-specified entire column-density probability distributions ($A_{\rm V}$-PDFs) to parametrize the ISM at large scales, linking them with a grid of pre-calculated one-dimensional PDR calculations of uniform-density distributions under different ISM environmental parameters. Here, we continue the above efforts by exploring in more detail the connection of $A_{\rm V}$-PDFs with the PDR models and by performing radiative transfer to calculate the emission of cooling lines from these distributions. In particular, we present {\sc PDFchem}\footnote{https://github.com/tbisbas/PDFchem}, a new fast numerical method that is able to compute the PDR chemistry of large-scale inhomogeneous ISM regions using probability distributions of physical parameters as an input within seconds.

As will be discussed in detail later, {\sc PDFchem} should be treated as an attempt to overcome the high computational cost needed by detailed three-dimensional hydrochemical simulations. As such, it should be considered as a machinery to understand the \emph{trends} in estimating the atomic and molecular gas content of large-scale ISM regions and to constrain the ISM environmental parameters which may be used for further in-depth investigations. The results presented here may vary if different initial conditions in the PDR calculations are used or if PDR simulations using other astrochemical codes are applied, e.g. the {\sc kosma}-$\tau$ code \citep{Stoe96,Roel22}, the {\sc meudon} code \citep{LePe06}, the {\sc Cloudy} code \citep{Ferl98} and others. The latter can be further understood given the very non-linear nature of PDR chemistry that occurs \citep[see e.g. the code comparison and benchmarking of][]{Roll07}. Limitations and approximations in the present approach include assumptions regarding the structure of the FUV radiation field, the neglect of a diffuse component of radiation, as well as the assumption that the cosmic-ray ionization and the metallicity are everywhere constant for the given $A_{\rm V}$-PDF. Section~\ref{ssec:limits} is dedicated to stress the importance of the above issues, paving the way for the development of future {\sc PDFchem} versions.

This paper is organized as follows. 
Section~\ref{sec:aveffavobs} defines the `effective' and `observed' visual extinctions used in this work.
Section~\ref{sec:methods} describes how the {\sc PDFchem} algorithm operates using the aforementioned effective and observed $A_{\rm V}$, compares to previous works and full hydro models and discusses the limitations of the method.
Section~\ref{sec:preamble} is a preamble to application results and describes the oxygen and carbon chemistry as well as the line ratios used in our application results. 
Section~\ref{sec:applications} presents results for two hypothetical $A_{\rm V}$-PDFs considered corresponding to a non-star-forming and a star-forming ISM distribution.
In Section~\ref{sec:discussion} we discuss the use of {\sc PDFchem} in high-redshift galaxies. We conclude in Section~\ref{sec:conclusions}.

\section{Effective and observed visual extinctions}
\label{sec:aveffavobs}

Consider a three-dimensional cloud consisting of a number of computational cells. Each cell has a total H-nucleus number density of $n_{\rm H}$. From each cell there are $\cal N_{\ell}$ directions emanating uniformly and distributed over the $4\pi$ celestial sphere, along which the local column density (and hence the visual extinction) is calculated. Throughout this work, we distinguish between two different $A_{\rm V}$ quantities: the `effective' (or `shielding' or `local' or `3D') visual extinction denoted as $A_{\rm V,eff}$, and the `observed' (or `2D') visual extinction denoted ad $A_{\rm V,obs}$. 

\begin{figure}
    \centering
	\includegraphics[width=0.22\textwidth]{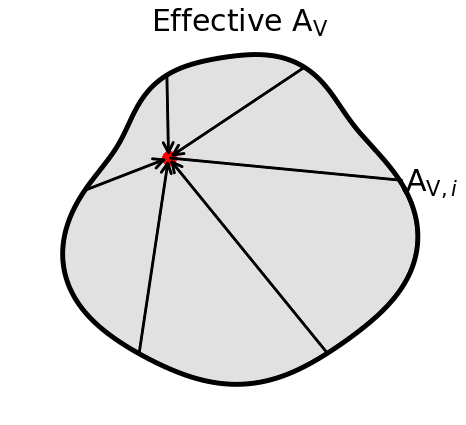}
	\includegraphics[width=0.24\textwidth]{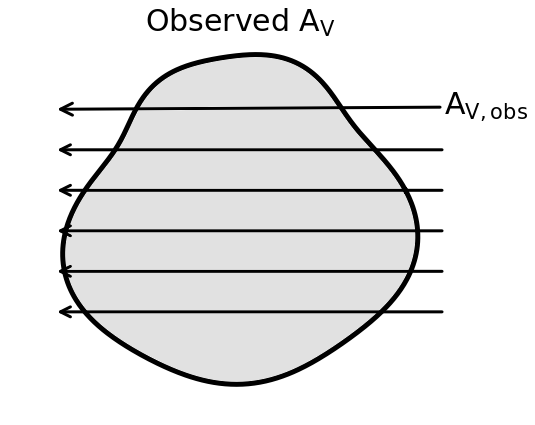}
    \caption{
    Schematic illustrating the differences between the effective (left) and the observed (right) visual extinctions in a hypothetical cloud. In the left case, the effective $A_{\rm V}$ is calculated as the average visual extinction of all $A_{{\rm V},i}$ that a cell (red dot) locally has (see Eqn.~\ref{eqn:aveff}) and along which the FUV radiation attenuates. This $A_{\rm V,eff}$ determines its PDR chemistry. In the right case, the $A_{\rm V,obs}$ is the visual extinction that a hypothetical observer measures as placed in the right hand side of the cloud.}
    \label{fig:cartoon}
\end{figure}

The $A_{\rm V,eff}$ represents the local visual extinction of a computational element consisting of the density distribution (see the left schematic of Fig.~\ref{fig:cartoon}). Under the assumption that the diffuse FUV component is negligible and that the radiation is always perpendicular to the surface, it is given by the expression \citep{Glov10,Offn13}:
\begin{eqnarray}
\label{eqn:aveff}
A_{\rm V,eff}=-\frac{1}{\gamma}\ln\left( \frac{1}{\cal N_{\ell}} \sum_{i=1}^{\cal N_{\ell}} e^{-\gamma A_{{\rm V},i}} \right),
\end{eqnarray}
where $\gamma=2.5$ is the dust attenuation factor \citep{Berg04,vDis08}, and $A_{{\rm V},i}$ is the visual extinction along each different direction on the celestial sphere centered on the computational element. In the {\sc 3d-pdr} code (see \S\ref{ssec:pdr}), these directions are controlled by the {\sc healpix} package \citep{Gors05} in the $\ell$ level of refinement. The above value of $\gamma$ refers to the dust shielding on CO, however our results will not be affected if a different $\gamma$ factor for H$_2$ or C is adopted \citep[e.g. a $\gamma=3.02$ as in][]{Roll07}. Along each of the aforementioned $A_{{\rm V},i}$ visual extinctions, the FUV radiation field is attenuated as well as the self-shielding of H$_2$ and CO molecules is computed \citep[see][for details]{Bisb12}. Consequently, the PDR chemistry of the given computational cell giving its abundances of species, gas and dust temperatures as well as level populations of coolants and their corresponding local emissivities, are all connected with this average, effective, visual extinction.

On the other hand, the $A_{\rm V,obs}$ represents the visual extinction along the line-of-sight of an external observer (see the right schematic of Fig.~\ref{fig:cartoon}). This is the actual column density that the various observations refer to via extinction mapping out of which the corresponding probability distributions ($A_{\rm V}$-PDFs) are constructed. While the line emission is a result of the local PDR chemistry as described above, the optical depth seen by the observer is connected with this, `observed' visual extinction. 

Hence, a connection between $A_{\rm V,obs}$ and $A_{\rm V,eff}$ is needed to convert the observed column densities to the corresponding effective ones and thus connect them with the local PDR chemistry.  In particular, both aforementioned $A_{\rm V}$ quantities are connected with the total H-nucleus number density. The ``$A_{\rm V,eff}-n_{\rm H}$ relation'' connects the $A_{\rm V,eff}$ with the local $n_{\rm H}$ (see \S\ref{ssec:avenh}). The ``$A_{\rm V,obs}-\langle n_{\rm H}\rangle$ relation'' connects the $A_{\rm V,obs}$ with the average, mass-weighted number density $\langle n_{\rm H}\rangle$ along the line of sight (see \S\ref{ssec:avonh}). In addition, the two $A_{\rm V}$ quantities are mutually connected via the ``$\langle A_{\rm V,eff}\rangle-A_{\rm V,obs}$ relation'' where $\langle A_{\rm V,eff}\rangle$ is, like before, the average mass-weighted effective visual extinction (see \S\ref{ssec:aveavo}). For simplicity, in this work it is assumed that the total H-nucleus number density, $\langle n_{\rm H}\rangle \equiv n_{\rm H}$ and that the effective visual extinction, $\langle A_{\rm V,eff}\rangle \equiv A_{\rm V,eff}$. It is noted here that the latter two assumptions do not imply a uniform density slab. Instead, these assumptions connect the $A_{\rm V,eff}$ with the local $n_{\rm H}$ as if the spread around the mean is negligible, i.e. a one-to-one correlation. Although the scatter can be large especially for low densities, it is these mean values that our approach adopts. As described below, this correlation leads to a one-dimensional variable density distribution (see Appendix~\ref{app:slab}) covering densities in the range of $n_{\rm H}=10^{-1}-10^6\,{\rm cm}^{-3}$. Such a one-dimensional distribution was also constructed by \citet{Hu21} which reproduced the PDR abundances of their three-dimensional hydrodynamical models.

\section{The {\sc PDFchem} algorithm}
\label{sec:methods}

\subsection{Overview of the method}
\label{ssec:overview}

The {\sc PDFchem} algorithm follows the methodology described in \citetalias{Bisb19} in which a set of pre-run PDR thermochemical calculations is used to compute the average abundances of species on cloud scales spanning tens-to-hundreds of pc. This methodology is extended here to compute the average antenna temperatures and to provide estimates for the most commonly used line ratios by solving the radiative transfer equation accordingly. The equation:
\begin{eqnarray}
\label{eqn:fsp}
f_{\rm sp} = \frac{\sum_{i=1}^{q} N_i({\rm sp})\times{\rm PDF}(A_{{\rm V},i})\Delta A_{{\rm V},i}}{\sum_{i=1}^{q} N_i({\rm H, tot})\times{\rm PDF}(A_{{\rm V},i})\Delta A_{{\rm V},i}}
\end{eqnarray}
was introduced in \citetalias{Bisb19} to calculate the average column-integrated fractional abundance, $f_{\rm sp}$, of a species (`sp'). In the above, $q$ represents the resolution in which the $A_{\rm V}-$PDF relation is divided to and $N_i$ is the column density multiplied by the frequency ${\rm PDF}(A_{{\rm V},i})\Delta A_{{\rm V},i}$. In practice, the required resolution is such that the PDF can be sampled with a few thousand clouds at different $A_{\rm V,obs}$. Using the above equation, estimates of the average gas temperature are also considered in this work.

Contrary to the use of several uniform-density slabs in \citetalias{Bisb19}, the PDR database in this work is constructed using a single slab with a density distribution as described below. This distribution is interacting with a large set of ISM environmental parameter combinations (\S\ref{ssec:pdr}) building the database of pre-run PDR simulations. The density distribution (hereafter `variable density slab') is constructed from the empirical $A_{\rm V,eff}-n_{\rm H}$ relation (\S\ref{ssec:avenh} and Appendix~\ref{app:slab}). With the above in mind, the following steps take place.

\begin{enumerate}
    \item The user inputs an $A_{\rm V,obs}-$PDF distribution representing the observed/simulated object.
    \item For each of the given $A_{\rm V,obs}$ values, {\sc PDFchem} finds the most probable $n_{\rm H}$ number density (\S\ref{ssec:avonh}).
    \item The above $n_{\rm H}$ corresponds to an effective position, $r$, in the density distribution of the variable slab and thus to an $A_{\rm V,eff}$. In turn, the column densities of species and the antenna temperatures of emission lines (radiative transfer) are calculated from the edge of the slab until the effective location $r$. 
    \item Steps (ii) and (iii) are repeated for every $A_{\rm V,obs}$, followed by application of Eqn.(\ref{eqn:fsp}) to estimate the final weighted-average PDR quantities (abundances of species, line emission and gas temperatures).
\end{enumerate}

The above steps are performed for all different combinations of ISM environmental parameters considered (\S\ref{ssec:pdr}). Here, a major assumption is that the overall ISM structure, as presented by the $A_{\rm V,obs}$-PDF distribution, is not affected by different parameters (see also \S\ref{ssec:limits}). Eventually, contour-maps are constructed showing the response of the above PDR quantities under the ISM combinations. In this work, these contour-maps examine the atomic-to-molecular transition, the abundances of the carbon cycle and other important species, the average gas temperature and the most frequently used cooling line ratios for all possible pairs of cosmic-ray ionization rate and FUV intensity considered at constant metallicity, $Z$.

\subsection{Photodissociation region chemistry}
\label{ssec:pdr}

We use the publicly available code {\sc 3d-pdr}\footnote{https://uclchem.github.io/3dpdr.html} \citep{Bisb12} which treats the astrochemistry of PDRs by balancing various heating and cooling processes as a function of column density. Once the code reaches equilibrium in which the total heating and total cooling are equal to within a user-defined tolerance parameter, it outputs the gas and dust temperature profile, the abundances of species, the level populations of the coolants as well as the various heating and cooling functions versus the column depth. We adopt the modifications and updates presented in \citetalias{Bisb19} which include the suprathermal formation of CO via CH$^+$ \citep{Fede96,Viss09}. 

In this CO formation path, as Alfv\'enic waves enter the cloud and dissipate as a function of depth, they cause non-thermal motions between ions and neutrals. This effect allows the reaction C$^+$+H$_2\rightarrow$~CH$^+$+H to overcome its energy barrier and thus form CH$^+$ which in turn reacts with O enhancing the CO abundance \citep{Fede96, Viss09}. The above reaction increases the CO abundance at low columns of H$_2$ and is in better agreement with observations \citep[][]{Rach02,Shef08}. We assume that the above route of CO formation occurs until $A_{\rm V}=0.7\,{\rm mag}$ from which point onward it ceases to be important. Recently, \citet{Hu21} found that time-dependency of H$_2$ formation plays an equally important and independent role with the suprathermal formation route of CO which can also explain the aforementioned observations. Interestingly, \citet{Sun22} found that in one-dimensional PDR simulations of collapsing clouds that have purely atomic hydrogen as initial chemical composition, the H + O $\rightarrow$ OH reaction is accelerated and CO may form through the OH channel \citep{Bisb17}. This also increases CO abundances at low column densities where H$_2$ is not the dominant hydrogen phase. These reactions are included in our network.

As in \citetalias{Bisb19}, a chemical network of 33 species and 330 reactions is used which is a subset of the UMIST2012 \citep{McEl13} network. Unless stated otherwise in the tests performed below, Table~\ref{tab:abun} shows the initial gas-phase elemental abundances of species used here, assuming a gas-to-dust ratio of 100. The column density of dust grains is taken to scale linearly with the metallicity, $Z$ \citep[e.g.][for values down to $Z=0.1\,{\rm Z}_{\odot}$]{Lero11,Herr12,Feld12}, thus the visual extinction, $A_{\rm V}$, also scales linearly with $Z$. 

A variety of different combinations of ISM environmental parameters between the incident FUV radiation field, $\chi/\chi_0$ \citep[normalized to the spectral shape of][]{Drai78}, the cosmic-ray ionization rate, $\zeta_{\rm CR}$ and the metallicity, $Z$, is used in this work. In particular, we consider 40 different FUV intensities in the range of $\chi/\chi_0=10^{-1}-10^3$ logarithmically spaced. The FUV intensity is attenuated as a function of depth along the line of sight \citep{Bisb12}. We also consider 40 different cosmic-ray ionization rates in the range of $\zeta_{\rm CR}=10^{-17}-10^{-13}\,{\rm s}^{-1}$ per H$_2$ logarithmically spaced. This parameter is kept constant throughout the cloud \citep[c.f.][]{Gach19}. These combinations of $\chi$ and $\zeta_{\rm CR}$ make a grid of 1,600 simulations for a constant metallicity. We also consider four different metallicities ($Z/Z_{\odot}=$ 0.1, 0.5, 1, 2), thus making a total of 6,400 PDR simulations. All these simulations use a single density distribution (the variable density slab) resulting from the $A_{\rm V,eff}-n_{\rm H}$ relation which covers a density range between $n_{\rm H}=10^{-1}-10^6\,{\rm cm}^{-3}$. It is further assumed that the slope of the variable density slab does not change as a function of metallicity \citep[c.f.][]{Hu21}. The construction of this slab is discussed below and in Appendix~\ref{app:slab}. The described set of PDR calculations constitutes the pre-run thermochemical database of the algorithm.

\begin{table}
	\centering
	\caption{Initial gas-phase abundance of species used for the grid of PDR simulations. The gas-to-dust ratio is assumed to be 100.}
	\label{tab:abun}
	\begin{tabular}{lcl}
		\hline
		\hline
		Element & Abundance & Reference \\
		\hline
		H     & $4.00\times10^{-1}$ & --\\
		H$_2$ & $3.00\times10^{-1}$ & --\\
		He    & $1.00\times10^{-1}$ & --\\
		C$^+$ & $1.40\times10^{-4}$ & \citet{Card96} \\
		O & $2.80\times10^{-4}$ & \citet{Cart04} \\
		\hline
	\end{tabular}
\end{table}

\subsection{The $A_{\rm V,eff}-n_{\rm H}$ relation and the variable density slab}
\label{ssec:avenh}

\begin{figure}
    \centering
	\includegraphics[width=0.48\textwidth]{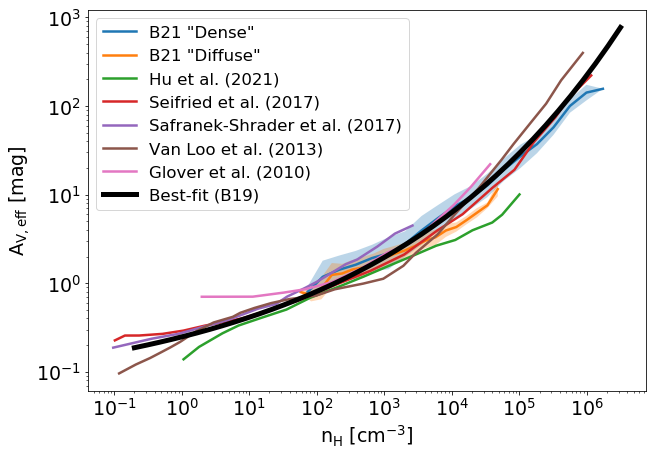}
	\includegraphics[width=0.48\textwidth]{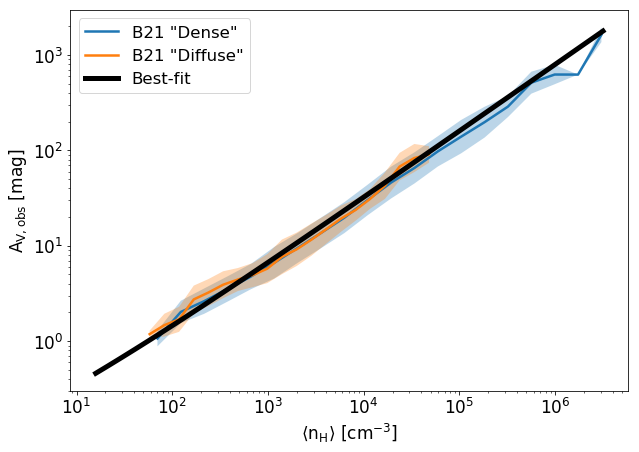}
	\includegraphics[width=0.48\textwidth]{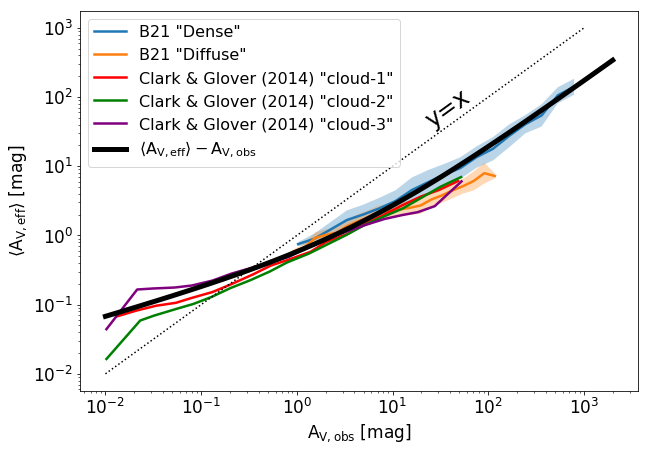}
    \caption{\textit{Top panel:} The $A_\mathrm{V,eff}{-}n_\mathrm{H}$ relation for the `Dense' (blue solid line) and `Diffuse' (orange solid line) clouds presented in \citetalias{Bisb21}. Results from the studies of \citet{Glov10, VanL13, Safr17, Seif17, Hu21} are also plotted for comparison. The $A_\mathrm{V,eff}$ best-fit in solid black line corresponds to Eqn.~\ref{eq:aveff_nh}. \textit{Middle panel:} The $A_\mathrm{V,obs}{-}\langle n_\mathrm{H}\rangle$ relation as calculated for the \citetalias{Bisb21} clouds. The black solid line corresponds to Eqn.~\ref{eq:avobs_nh}. \textit{Bottom panel:} The $\langle A_\mathrm{V,eff}\rangle{-}A_\mathrm{V,obs}$ relation calculated for these clouds in black solid line. The red, green and purple coloured lines correspond to the mean values of three molecular clouds simulated in \citet{Clar14}. The dotted line in this panel corresponds to the 1-1 relation ($y=x$) to guide the eye. In all panels the shaded regions correspond to $1\sigma$ deviation in the \citetalias{Bisb21} simulations.}
    \label{fig:avnh}
\end{figure}

In \citetalias{Bisb19}, a relation between the $A_{\rm V,eff}$ and the local number density, $n_{\rm H}$, was presented which gives the most probable value of the effective visual extinction for a given $n_{\rm H}$. It results from a collection of four different simulations \citep[from pc to kpc scales,][]{Glov10, VanL13, Safr17, Seif17, Hu21}. The best-fit equation describing it is
\begin{eqnarray}
A_{\rm V,eff}(n_{\rm H}) = 0.05 \exp\left\{{1.6\left(\frac{n_\mathrm{H}}{[\mathrm{cm}^{-3}]}\right)^{0.12}}\right\}\,[\mathrm{mag}].
\label{eq:aveff_nh}
\end{eqnarray}

The upper panel of Fig.~\ref{fig:avnh} shows Eqn.~\ref{eq:aveff_nh} in black solid line, along with the results from the aforementioned simulations for comparison. In addition, the $A_{\rm V,eff}{-}n_{\rm H}$ relation for the two three-dimensional MHD clouds studied in \citetalias{Bisb21} and post-processed with {\sc 3d-pdr}, are plotted in blue (the `Dense' cloud) and in orange (the `Diffuse' cloud) solid lines while the corresponding shaded region is the $1\sigma$ standard deviation in the \citetalias{Bisb21} models. These two relations are calculated using ${\cal N_{\ell}}=12$ {\sc healpix} rays \citep{Gors05} (see Eqn.~\ref{eqn:aveff}).

Following the $A_\mathrm{V,eff}{-}n_\mathrm{H}$ relation described in Eqn.~\ref{eq:aveff_nh}, the one-dimensional variable density distribution is constructed which will be used to compute the grid of astrochemical simulations for the various ISM parameters (see Appendix~\ref{app:slab}). Contrary to the standard way of building a grid of uniform density one-dimensional slabs and calculating the astrochemical properties under different ISM environmental parameters (e.g. as described in \citetalias{Bisb19}), throughout the present work the variable density distribution resulting from the $A_{\rm V,eff}{-}n_{\rm H}$ relation -and only this- will be used.

\subsection{The $A_{\rm V,obs}{-}\langle n_{\rm H}\rangle$ relation}
\label{ssec:avonh}

While $A_{\rm V,eff}$ is the local visual extinction in the three-dimensional density distribution, the projected (observed) visual extinction is derived from the column density of the gas that is along the line-of-sight of the observer. We connect this observed visual extinction, $A_{\rm V,obs}$, with the projected number density which is the average density along the line-of-sight. The latter is calculated from the three-dimensional distributions following the mass-weighted relation
\begin{eqnarray}
\langle n_{\rm H}\rangle = \frac{\int n_{\rm H}m_{\rm cell} dr}{\int m_{\rm cell} dr},
\end{eqnarray}
where $m_{\rm cell}$ is the mass of the computational grid cell. 

The middle panel of Fig.~\ref{fig:avnh} shows the $A_{\rm V,obs}{-}\langle n_{\rm H}\rangle$ correlation. Both modeled clouds of \citetalias{Bisb21} are in excellent agreement with each other albeit that the `Dense' cloud contains densities up to two orders of magnitude higher than the `Diffuse' cloud. The resultant correlation is parametrized with the relation
\begin{eqnarray}
\label{eq:avobs_nh}
A_\mathrm{V,obs} = 0.06 \left(\frac{\langle n_\mathrm{H}\rangle}{[\mathrm{cm^{-3}}]}\right)^{0.69}\mathrm{[mag]},
\end{eqnarray}
and is illustrated in black solid line in the aforementioned panel. Unlike with the $A_{\rm V,eff}{-}n_{\rm H}$ relation, the $A_{\rm V,obs}{-}\langle n_{\rm H}\rangle$ one appears to be a simple power-law, at least to the extent explored in the present work. 
Following the method for the variable density profile using the $A_{\rm V,eff}-n_{\rm H}$ relation, the corresponding one derived from the $A_{\rm V,obs}{-}\langle n_{\rm H}\rangle$ is also obtained (see Appendix~\ref{app:slab}).

\subsection{The $\langle A_{\rm V,eff}\rangle{-}A_{\rm V,obs}$ relation}
\label{ssec:aveavo}

Considering the previous $A_{\rm V,eff}{-}n_{\rm H}$ and $A_{\rm V,obs}{-}\langle n_{\rm H}\rangle$ relationships, the $\langle A_{\rm V,eff}\rangle{-}A_{\rm V,obs}$ can be readily extracted. In particular, the combination of Eqns.~\ref{eq:aveff_nh} and~\ref{eq:avobs_nh} results in:
\begin{eqnarray}
\langle A_{\rm V,eff}\rangle = 0.05\exp\left\{2.6\left(\frac{A_{\rm V,obs}}{[{\rm mag}]}\right)^{0.17}\right\}.
\label{eq:aveff_avobs}
\end{eqnarray}
It is interesting, however, to explore how the latter relation compares with results from hydrodynamical simulations. Such an $\langle A_{\rm V,eff}\rangle{-}A_{\rm V,obs}$ relationship was discussed in \citet{Clar14} while studying the star formation rate of three different molecular clouds\footnote{We denote as `cloud-1' the `n100\_m1250', as `cloud-2' the `n26\_m10000', and as `cloud-3' the `n264\_m10000' simulations of \citet{Clar14}.}. The results of those calculations are illustrated with red, green and purple coloured lines in the bottom panel of Fig.~\ref{fig:avnh}, which cover a range of $A_{\rm V,obs}{\sim}10^{-2}{-}50\,{\rm mag}$. The black solid line represents Eqn.~\ref{eq:aveff_avobs}.

To correlate the $A_{\rm V,eff}$ and $A_{\rm V,obs}$ from the 3D hydrodynamical clouds of \citetalias{Bisb21}, we account for the mass-weighted $A_{\rm V,eff}$ calculated as
\begin{eqnarray}
\langle A_{\rm V,eff}\rangle = \frac{\int A_{\rm V,eff}m_{\rm cell}dr}{\int m_{\rm cell}dr}.
\end{eqnarray}
As with the previous cases, the resulting correlation is illustrated in blue and orange solid lines in the bottom panel of Fig.~\ref{fig:avnh}. While both `Dense' and `Diffuse' clouds are more massive and compact from those modeled by \citet{Clar14}, they cover a range of $A_{\rm V,obs}{\sim}10^0{-}8\times10^2\,{\rm mag}$. As can be seen in the bottom panel of Fig.~\ref{fig:avnh}, Eqn.~\ref{eq:aveff_avobs} is in agreement with both suites of \citet{Clar14} and \citetalias{Bisb21} hydrodynamical clouds. 

In classical PDR studies, it is frequently assumed that $A_{\rm V,eff}\equiv A_{\rm V,obs}$. This case is plotted with dotted lines in the bottom panel of Fig.~\ref{fig:avnh}. As can be seen, for $A_{\rm V,obs}>1\,{\rm mag}$, the $A_{\rm V,obs}$ is larger than $A_{\rm V,eff}$ and the difference becomes substantial, thus a connection of the two visual extinctions, such as Eqn.~\ref{eq:aveff_avobs}, must be adopted for a proper explanation of the observed clouds.

\subsection{Line emission}
\label{ssec:rt}

To calculate the emission of each cooling line as seen by the observer for a given transition $i\rightarrow j$, the equation of radiative transfer
\begin{eqnarray}
\label{eqn:RT}
\frac{dI_{\nu}}{dz}=-\alpha_{\nu}I_{\nu}+\alpha_{\nu}S_{\nu},
\end{eqnarray}
must be solved along the line-of-sight of the observer ($z$-axis). The line emission is calculated using the level populations outputted by {\sc 3d-pdr}, after thermal balance has been reached. In a given $A_{\rm V,obs}-$PDF, this equation is solved for all $A_{\rm V,obs}$, following the relations discussed in \S\ref{ssec:avenh}-\ref{ssec:aveavo}. In the above equation,
\begin{eqnarray}
\alpha_{\nu}=\frac{c^2n_iA_{ij}} {8\pi\nu_0^2} \left(\frac{n_jg_i}{n_ig_j}-1\right)\phi_{\nu},
\end{eqnarray}
is the absorption coefficient and 
\begin{eqnarray}
S_{\nu}=\frac{2h\nu_0^3}{c^2}\frac{n_ig_j}{n_jg_i-n_ig_j},
\end{eqnarray}
is the source function at frequency $\nu$. In the above, $A_{ij}$ is the Einstein A coefficient, $n_i$ and $n_j$ are the level populations, $g_i$ and $g_j$ the corresponding statistical weights and $\phi_{\nu}$ is the line profile. During the thermal balance iterations, {\sc 3d-pdr} calculates the level populations of the cooling lines adopting the Large Velocity Gradient (LVG; \citealt{Sobo60}) escape probability approach in which the thermal line width is much smaller than the line width due to the significant line-of-sight velocity change in each direction. Then a photon at frequency $\nu_{ij}$ escapes from the modelled cloud without interacting with its surrounding ISM gas with a probability:
\begin{eqnarray}
    \beta_{ij} = \frac{1-e^{-\tau_{ij}}}{\tau_{ij}},
\end{eqnarray}
where $\tau_{ij}$ is the optical depth for the corresponding frequency.

The line profile, $\phi_{\nu}$, for a non-moving gas along the line-of-sight of the observer is given by the expression:
\begin{eqnarray}
\phi_{\nu} = \frac{1}{\sqrt{2\pi\sigma_{\nu}^2}},
\end{eqnarray}
where $\sigma_{\nu}$ is the dispersion (which is one half of the Doppler width) defined as
\begin{eqnarray}
\sigma_{\nu} = \frac{\nu_{ij}}{c}\sqrt{\frac{k_{\rm B}T_{\rm gas}}{m_{\rm mol}} + \frac{v_{\rm turb}^2}{2}}.
\end{eqnarray}
In the above, $T_{\rm gas}$ is the local gas temperature, $m_{\rm mol}$ is the molecular weight of the emitting coolant and $v_{\rm turb}$ is the root-mean-square velocity dispersion due to microturbulence, which will be treated as a constant (and taken to be $v_{\rm turb}=1\,{\rm km}\,{\rm s}^{-1}$ unless otherwise stated) in the present work. 

The radiative transfer Eqn.~(\ref{eqn:RT}) is then numerically solved following the methodology of \citet{Bisb17b} (see also \citetalias{Bisb21} for an update to include treatment of dust opacity which is also used here). Once solved, the intensity $I_{\nu}$ is converted to the antenna temperature using the relation
\begin{eqnarray}
T_{\rm A} = \frac{c^2 I_{\nu}}{2 k_{\rm B} \nu_{ij}^2}\ [{\rm K}],
\end{eqnarray}
and by multiplying the above quantity with the linewidth, the velocity integrated emission, $W$ [$\rm K\,km\,s^{-1}$], is obtained. Note that the radiative transfer scheme presented above and used in {\sc PDFchem} has been benchmarked against {\sc radex} \citep{vdTak07} with the results presented in Appendix~\ref{app:radex}.

To calculate the average antenna temperature from a given distribution of $A_{\rm V,obs}$, we apply a modified version of Eqn~\ref{eqn:fsp} in the form
\begin{eqnarray}
f(T_{\rm A})=\frac{\sum_{i=1}^{q} T_{{\rm A},i}\times{\rm PDF}(A_{{\rm V},i})\Delta A_{{\rm V},i}}{\sum_{i=1}^{q} {\rm PDF}(A_{{\rm V},i})\Delta A_{{\rm V},i}}. 
\end{eqnarray}
Similarly, the average values of optical depths (discussed in Appendices~\ref{sssec:tau_nSF} and \ref{sssec:tau_SF}) are calculated.

\subsection{Comparison with the \citetalias{Bisb21} hydro models}

\begin{figure}
    \centering
	\includegraphics[width=0.95\linewidth]{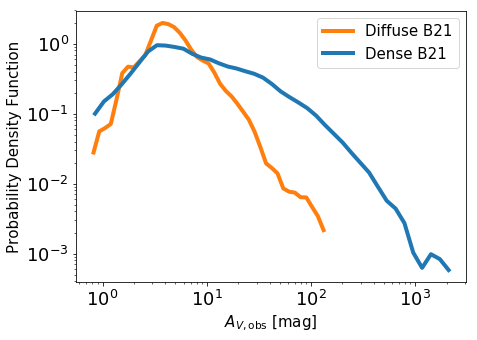}
    \caption{The $A_{\rm V,obs}$-PDF relations for the `Diffuse' (orange line) and the `Dense' (blue line) clouds of the 3D MHD simulation by \citetalias{Bisb21}.}
    \label{fig:avpdfB21}
\end{figure}

To benchmark and validate {\sc PDFchem} in using the one-dimensional variable density slab discussed above instead of a large grid of uniform density slabs, we compare it with the results of \citetalias{Bisb21} who studied two different three-dimensional MHD distributions of the same size ($14\,{\rm pc}$) and post-processed with {\sc 3d-pdr}.

Figure~\ref{fig:avpdfB21} shows the $A_{\rm V,obs}$-PDFs of the two MHD clouds. Both functions peak at ${\sim}4\,{\rm mag}$ ({\it mode} of the PDF; see \citetalias{Bisb19}), however the `Dense' PDF has a larger width than the PDF of the `Diffuse' cloud and therefore higher values of $A_{\rm V,obs}$ are considered. Using these two probability functions, we shall estimate the average values of the abundances of species and ratios of emission lines. 

We compare the results for the abundances under different ISM environmental parameters as well as for the radiative transfer calculations, using {\sc PDFchem}. Following the chemistry adopted in \citetalias{Bisb21}, we switch off the suprathermal formation of CO via CH$^+$ for this comparison. We also consider a miroturbulent velocity of $v_{\rm turb}=2\,{\rm km}\,{\rm s}^{-1}$.

\begin{figure}
    \centering
	\includegraphics[width=0.425\textwidth]{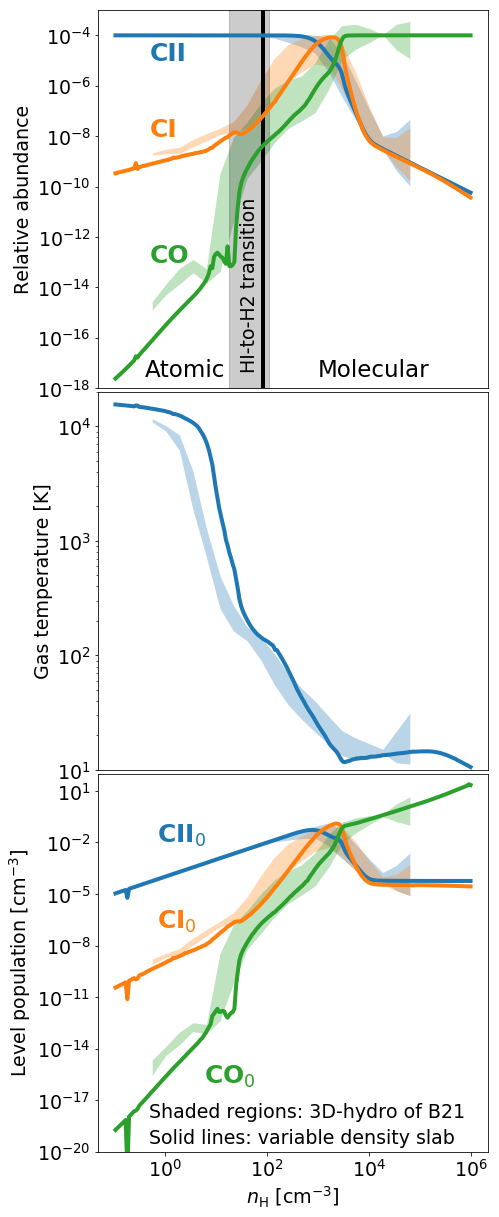}
    \caption{Comparison between a three-dimensional MHD simulation with PDR post-processing (shaded regions representing $1\sigma$ standard deviation) and a PDR calculation using the variable density slab (solid lines). The three-dimensional results are taken from \citetalias{Bisb21} (`Diffuse' cloud) for ISM conditions corresponding to $\zeta_{\rm CR}=10^{-16}\,{\rm s}^{-1}$, $\chi/\chi_0=10$ and $Z=1\,{\rm Z}_{\odot}$. The top panel compares the abundances of C{\sc ii} (blue), C{\sc i} (orange) and CO (green) as a function of the local number density, $n_{\rm H}$. The gray region shows the H{\sc i}-to-H$_2$ transition in the three-dimensional simulation and it corresponds to the condition $|x({\rm HI})-2x({\rm H_2})|<10^{-3}$, while the black solid line is the aforementioned transition in the one-dimensional simulation. The middle panel shows the gas-temperature as a function of $n_{\rm H}$. The bottom panel shows the level population number density of C{\sc ii}, C{\sc i} and CO in the ground level. The agreement is seen to be very good.}
    \label{fig:3Dvs1D_abund}
\end{figure}

Figure~\ref{fig:3Dvs1D_abund} shows a comparison between the `Diffuse' cloud of \citetalias{Bisb21} (shaded regions representing $1\sigma$ standard deviation around the mean) and results obtained using the variable density slab (solid lines). The ISM environmental conditions considered are $\zeta_{\rm CR}=10^{-16}\,{\rm s}^{-1}$, $\chi/\chi_0=10$ FUV intensity and $Z=1\,{\rm Z}_{\odot}$ metallicity, which correspond to the fiducial ISM parameters of \citetalias{Bisb21}. The gray region corresponds to the H{\sc i}-to-H$_2$ region which satisfies the abundance difference condition of $|x({\rm HI})-2x({\rm H_2})|<10^{-3}$. Solid lines correspond to the results of the one-dimensional simulation. We compare the abundances of C{\sc ii}, C{\sc i} and CO, the gas temperature and the level population number density of the aforementioned species for the ground level as a function of the total number density, $n_{\rm H}$. The latter two quantities are particularly important for the radiative transfer calculations. As can be seen, in all cases the variable density slab reproduces very well the computationally expensive three-dimensional models. After experimenting, this agreement is true for all ISM environmental parameters explored in \citetalias{Bisb21} ($\zeta_{\rm CR}=10^{-17}-10^{-14}\,{\rm s}^{-1}$, $\chi/\chi_0=1-10^3$ and $Z=0.1-2\,{\rm Z}_{\odot}$).

\begin{figure}
    \centering
	\includegraphics[width=0.99\linewidth]{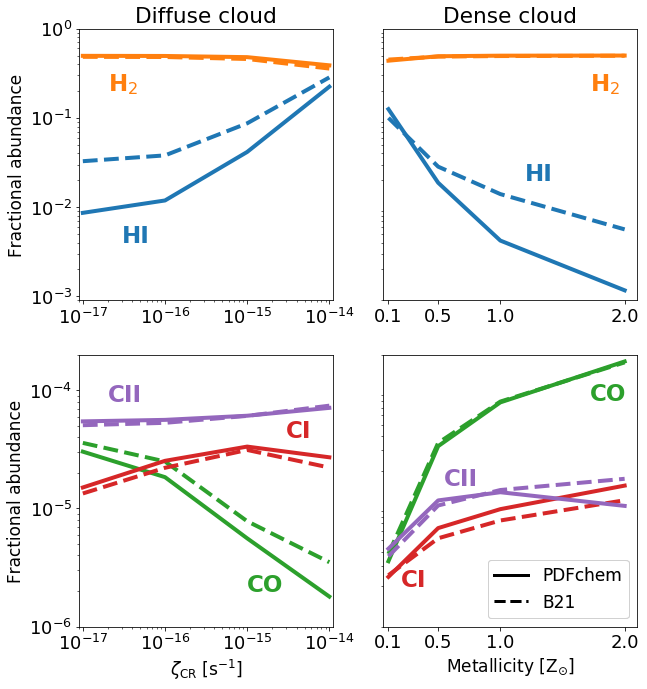}
    \caption{Comparison with \citetalias{Bisb21} MHD simulations. Left column shows the `Diffuse cloud' comparison with the cosmic-ray ionization rate as the free parameter. Right column shows the `Dense cloud' comparison with the metallicity as the free parameter. Top row shows the atomic-to-molecular transition and bottom row the carbon cycle. Solid lines show the results using {\sc PDFchem} and dashed lines those of \citetalias{Bisb21}. As can be seen, the results are in a broad agreement.}
    \label{fig:vB21}
\end{figure}

Figure~\ref{fig:vB21} shows the comparison between the abundances of \hi, H$_2$ (top row) and the carbon cycle (bottom row) as calculated from the 3D MHD simulations of \citetalias{Bisb21} versus those calculated using {\sc PDFchem} and after applying it to the entire $A_{\rm V,obs}$-PDFs of the two clouds. For the `Diffuse' cloud comparison (left column), we use the cosmic-ray ionization rate as the free parameter while keeping the FUV intensity fixed at $\chi/\chi_0=10$ and at solar metallicity. For the `Dense' cloud (right column) we use the metallicity as the free parameter while keeping the cosmic-ray ionization rate fixed at $\zeta_{\rm CR}=10^{-16}\,{\rm s}^{-1}$ and using the aforementioned FUV intensity. As can be seen, we obtain good agreement between {\sc PDFchem} and the 3D MHD simulations. In particular, the `Diffuse' cloud remains always molecular and C{\sc ii}-dominated for any $\zeta_{\rm CR}=10^{-17}-10^{-14}\,{\rm s}^{-1}$. The `Dense' cloud also remains molecular but CO-dominated for any value of $Z=0.1-2\,{\rm Z}_{\odot}$. 

\begin{figure*}
    \centering
	\includegraphics[width=0.98\linewidth]{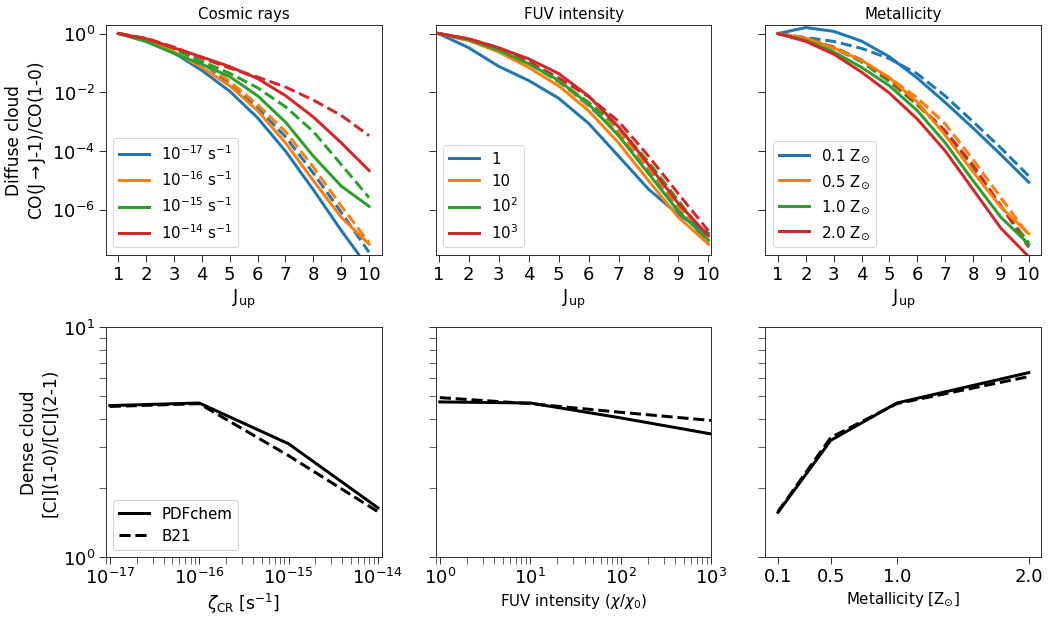}
    \caption{Comparison with the radiative transfer results of the 3D MHD models of \citetalias{Bisb21}. Top row shows the CO SLEDs for the `Diffuse' cloud, normalized to the $J=1{-}0$ transition. Bottom row shows the emission ratio for the atomic carbon lines ([\ci]($1{-}0$)/[\ci]($2{-}1$)). Columns from left to right show the results for each different free parameter considered (cosmic-ray ionization rate, FUV intensity, metallicity). Solid lines show results using {\sc PDFchem} and dashed lines the \citetalias{Bisb21} results. As can be seen, we achieve very good agreement with the 3D results which dramatically decreases the computational cost in examining PDRs of complex structures.}
    \label{fig:vRT}
\end{figure*}

It is interesting also to explore the comparison how {\sc PDFchem} compares with the 3D MHD of \citetalias{Bisb21} runs for the radiative transfer calculations. Figure~\ref{fig:vRT} shows CO spectral line energy distributions (SLEDs) for the `Diffuse' cloud (top row) normalized to the $J=1{-}0$ transition and the atomic carbon line ratio ([\ci]($1{-}0$)/[\ci]($2{-}1$)) for the `Dense' cloud (bottom row). From left-to-right, we consider $\zeta_{\rm CR}$, the FUV intensity and the metallicity as the free parameter, respectively. In terms of the trends, we find excellent agreement in all cases and for both clouds. The {\sc PDFchem} and the 3D simulation only differ for higher $J$-transitions and for enhanced $\zeta_{\rm CR}$ values. Such differences may appear in extreme environments due to the sensitive dependence of the high-$J$ CO emission on the local gas temperature and density.

Finally, it is interesting to compare the computational cost between the two approaches. We find that, on top of an already expensive MHD simulation, the demanding three-dimensional density distributions require more than three orders of magnitude longer time to iterate over thermal balance and reach equilibrium, than the corresponding one-dimensional variable density distribution of Appendix~\ref{app:slab}. The small errors between the two approaches and the very high difference in computational time, makes {\sc PDFchem} an attractive new alternative to the complicated three-dimensional simulations for a quick exploration of how the different ISM environmental parameters may affect the PDR properties of given column density distributions. By assuming different PDF cases, it is also possible to skip the expensive hydrodynamical calculations and understand the behaviour of the atomic and molecular mass content under any combination of ISM environmental parameters.

\subsection{Limitations and applicability of {\sc PDFchem}}
\label{ssec:limits}

%The presented methodology and approach should be treated as an attempt to overcome the high computational cost needed by detailed three-dimensional hydrochemical simulations. As such, it should be considered as a machinery to understand the trends in estimating the atomic and molecular gas content of large-scale ISM regions and to constrain the ISM environmental parameters which may be used for further in-depth investigations. 

In order for {\sc PDFchem} to be applicable for the input column density distribution, the assumption that the ISM environmental parameters are \emph{symmetrical compared to the size of the modelled system} must be obeyed. This means that all environmental parameters shall either be treated in a radial manner (e.g. an isotropic FUV radiation field\footnote{Throughout this work, the isotropic radiation field is treated as a radial field without taking into account the contribution from the diffuse component due to scattering.}) or assumed to be constant everywhere in the distribution (e.g. cosmic-ray energy densities, metallicity).

Contributions to the diffuse FUV radiation field can arise from scattering of photons due to dust grains which can affect the penetration of FUV along a column \citep[e.g.][]{Flan80,LePe06,Roel13}. In particular, \citet{Goic07} showed that the PDR properties may change even more if a dust growth model along a column is assumed. However, earlier work by \citet{vanD84} showed that the photodissociation lifetime of OH varies by a factor of $\lesssim2$ for $A_{\rm V}\gtrsim1\,{\rm mag}$ between a forward scattering model and an isotropic one. For both C{\sc i} and CO, it is expected that the effect would be smaller since H$_2$ self-shielding dominates over dust shielding. Although photon scattering is an important FUV component for a more accurate PDR treatment, the presented results are not expected to significantly change. 

The diffuse component of radiation field can also arise from internal sources located mainly in the high density medium. For instance, feedback from massive stars such as H{\sc ii}-regions \citep[e.g.][]{Hu17,Oliv21} and supernova (SN) explosions \citep[e.g.][]{Walc15,Hopk18}, can significantly affect the distribution of the FUV radiation field and thus alter its isotropic assumption. SNe can further release large amounts of charged particles in the ISM changing its density distribution \citep{Giri16b} and increasing significantly $\zeta_{\rm CR}$. The latter is supported by observations suggesting an approximately ten percent conversion of the SN explosion energy to cosmic-ray energy \citep{Morl12,Held13}. On-going star-formation and in particular cluster formation can also increase locally the FUV radiation field \citep[see reviews by][]{Krum19,Rose20}. The cosmic-ray ionization rate may also change and largely vary in case protostellar objects are present \citep{Gach19}. While such effects can co-exist in $A_{\rm V,obs}-$PDFs of kpc-scale structures, it is assumed that their contribution to the global ISM environmental parameters is small. To take such local variations into account, one can in principle also adopt an FUV-PDF or a $\zeta_{\rm CR}$-PDF \citepalias[see also][]{Bisb19}.

Furthermore, a constant $\zeta_{\rm CR}$ with depth into the density distribution has been considered everywhere in this work.
However the low-energy component of cosmic-rays, which is responsible for the ignition of chemical reactions at high column densities, do attenuate as a result of Coulomb interactions and ionizations \citep{Pado09}. A treatment of cosmic-ray attenuation as a function of depth should be accounted for using a more realistic approach. It is noted, however, that recent three-dimensional simulations of Milky-Way clouds by \citet{Gach22} indicate that the constant $\zeta_{\rm CR}$ assumption may introduce only small relative errors to the global results for PDR tracers as opposed to a cosmic ray attenuated model. Similarly, a constant metallicity as well as a constant dust-to-gas ratio (which is linearly related to $Z$) has been considered in this work. In general, this assumption holds well for regions with extents up to kpc-scales, although metallicity gradients showing a decrease of $Z$ as a function of the galactocentric radius have been observed \citep[e.g.][]{Wolf03}.

The chemistry in all PDR simulations has been evolved up to a chemical time of $10\,{\rm Myr}$, at which point {\sc 3d-pdr} assumes that equilibrium has been reached \citep{Bisb12}. This is a reasonable time duration for the species we examine \citep[see][for a discussion]{Hold22}. However, time-dependency plays an important role in the H$_2$ formation particularly at low metallicities, as shown in hydrodynamical simulations by \citet{Hu21}. The steady-state assumption we adopt in this work may overestimate the H$_2$ abundance at low metallicities and in column-density distributions corresponding to a diffuse ISM region. 

In addition, we have assumed that the $A_{\rm V,eff}-n_{\rm H}$ relation -and thus the distribution of the variable density slab- remain the same under all ISM environmental parameters explored and that the spread in $A_{\rm V,eff}$ values for a constant $n_{\rm H}$ is negligible. The hydrodynamical models of \citet{Hu21} show that the effective column density may have a slightly different dependence on $n_{\rm H}$ for different metallicities. This in turn will impact the PDR chemistry and thus the final results. Thus for a more accurate modelling, a function relating the input $A_{\rm V,eff}-n_{\rm H}$ dependence with the various combinations of ISM environmental parameters need to be taken into account \citep[see also][]{Ploe20}. 

Finally, X-rays and shocks have not been considered in our PDR calculations. X-rays can affect the chemistry at high column densities through the production of secondary electrons (similarly to the way cosmic-rays do) which have kinetic energies high enough to cause ionizations of molecules and atoms \citep{Malo96,Meij05,Meij07,Mack19}. Shocks can also provide copious amounts of energy in the gas, which can in turn activate reactions with high energy barriers \citep{Meij11,Jame20} such as the production of OH through the reaction of O and H$_2$. Shocks are present in cloud-cloud collisions, in protostellar outflows, but also in galactic scale outflows. However, each of the aforementioned free parameters adds more complexity in the presented algorithm and dramatically increases the total number of PDR simulations needed for this treatment. Shocks do not dominate the ISM emission on large scales, except in case of mergers and AGN outflows \citep[e.g.][]{Meij13,Bell20}. Further developments of our algorithm may relax these constraints and allow to explore the trends as a function of these additional ISM parameters.

\section{Chemistry results: background}
\label{sec:preamble}

Before proceeding with presenting the results of our applications in the next Section (\S\ref{sec:applications}), it is sensible to provide a short introduction to the chemical abundances and the line ratios that will be explored. In particular, for each of the $A_{\rm V,obs}$-PDFs considered, the response of the atomic-to-molecular transition, the abundances of the carbon cycle \citepalias[discussed in Section 2 of][]{Bisb19}, the abundances of OH, OH$^+$, H$_2$O$^+$, CH, HCO$^+$ and the average (density-weighted) gas temperature will be studied under different combinations of ISM environmental parameters. By performing radiative transfer, the response of the most frequently used line ratios of the carbon cycle will be also studied. These include the atomic carbon line ratio ([C{\sc i}](2-1)/[C{\sc i}](1-0)), the first five rotational transitions of CO normalized to CO(1-0), and the line ratios of [C{\sc i}](2-1)/CO(7-6), [C{\sc i}](1-0)/CO(4-3), [C{\sc i}](1-0)/CO(1-0), [C{\sc ii}]/[C{\sc i}](1-0) and [C{\sc ii}]/CO(1-0). 

\subsection{Oxygen and carbon chemistry}

The molecule of OH can be formed through the reaction of O with H$_2$ or through the reaction of  H$_2$O$^+$ with H$_2$ followed by dissociative recombination with electrons of the produced H$_3$O$^+$ \citep{Bial15}. It can be also formed via the destruction of H$_2$O molecule with He$^+$ or H$^+$ as reactants, both sensitive to the $\zeta_{\rm CR}$ rate \citep{Meij11,Bisb15}. The latter routes are favoured in low-temperature regions. OH is an important molecule for the study of the ISM, as in regions permeated by high $\zeta_{\rm CR}$ values it can lead to the formation of CO \citep{Bisb17b}. Its line emission has been also suggested to trace the ``CO-dark'' \citep{vDis92} molecular gas \citep{Alle15,Li18b,Tang21}.

Cosmic-rays initiate the production of H$_3^+$ which in turn reacts with O forming OH$^+$. This formation route of OH$^+$ via proton transfer is mostly met in molecular regions where the gas temperature is low and the abundance of H$_2$ is high enough. In addition, cosmic rays reacting with H followed by the reaction with O enable the formation of O$^+$. H abstraction of the latter via its reaction with H$_2$ results in OH$^+$ \citep{Holl12,Bial15,Bial19}. At low column densities, OH$^+$ may also be formed via photoionization of OH, as well as through the reaction H$^+$+OH$\rightarrow$OH$^+$+H \citep{Meij11}. The abundance and thus the line emission of OH$^+$ is known to be very sensitive to the value of cosmic-ray ionization rate and it is thus frequently used to constrain $\zeta_{\rm CR}$ \citep{Holl12,Indr15}. The chemistry of H$_2$O$^+$ follows tightly that of OH and OH$^+$ species. H$_2$O$^+$ depends sensitively on the $\zeta_{\rm CR}$ value and it is also used as a tracer to constrain it \citep{Holl12,Indr15,Geri16,Neuf17}. 

The CH molecule has different formation routes depending on the density of the gas \citep{Blac73}. At low gas densities and thus low visual extinctions, CH is formed via dissociative recombination of CH$_2^+$, with the latter produced by radiative association of C$^++\,$H$_2$ \citep[e.g.][]{Gred93}. CH$_2^+$ then reacts with H$_2$ to produce CH$_3^+$. Both CH$_2^+$ and CH$_3^+$ products can lead to the formation of CH via dissociative recombination. At higher gas densities and thus higher visual extinctions the reaction of CH$_2$ with H produces CH and H$_2$. The observational work of \citet{Shef08} indicates that CH traces the H$_2$ that exists in the diffuse ISM. In the first extragalactic observational study of CH, \citet{Rang14} have used the CH/CO ratio to infer to the presence of X-rays in NGC~1068, and it is therefore an interesting molecule to study. 

The HCO$^+$ ion forms through the reaction of H$_3^+$ with CO. For this reaction to occur, the gas needs to be UV shielded -so for CO to form- and cosmic-rays must be present to ignite the formation paths leading to H$_3^+$. On the other hand, the destruction of HCO$^+$ occurs via reactions with H$_2$O and OH, as well as with recombination with electrons leading to the formation of CO. HCO$^+$ is considered to be a dense gas tracer \citep[e.g.][]{Knud07,Zhan14}. It is also frequently used in tandem with HCN due to the small difference in rest frequencies of their line emission, to infer to the star-formation efficiency, the dense gas properties or the X-ray contribution \citep[e.g.][]{Shim17,Gala20,Wolf22}.

\subsection{Line ratios}
\label{ssec:ratios}

Line ratios are frequently used as a diagnostic to infer the molecular gas content and to constrain the ISM environmental parameters. In this work, we focus on the most commonly used ratios between [C{\sc ii}]~$158\mu$m, the two [C{\sc i}] fine-structure lines and the first five CO transitions. 

The atomic carbon line ratio ([C{\sc i}](2-1)/[C{\sc i}](1-0), with rest frequencies of 809.34 and 492.16~GHz, respectively) is considered to constrain the average gas temperature \citep{Stut97,Schn03,Weis03,Papa04,Vale20} and it appears to be generally sub-thermally excited in the extragalactic context \citep{Papa22}. It has been used also to construct a distribution function of thermal pressures of ISM gas that best represents the Cold Neutral Medium of our Galaxy \citep{Jenk11}. The atomic carbon line ratio is of particular interest for the upcoming CCAT-prime telescope as its available frequency range will favor observations of this line ratio especially in our Galaxy. 

The very small (${\sim}2.7$~GHz) frequency difference of [C{\sc i}](2-1) and CO(7-6) allows to obtain these two lines simultaneously, thus their ratio is frequently explored, all the more since ALMA is able to capture it at high-redshift. In a similar fashion [C{\sc i}](1-0) and CO(4-3) can be observed simultaneously with ALMA for high-redshift systems and thus their ratio is also frequently used as a diagnostic of the molecular gas. [C{\sc i}](1-0) and CO(1-0) are major coolants of the dense star-forming gas. CO(1-0) is known to trace the total molecular mass \citep[see][for a review]{Bola13} while higher-$J$ transitions are useful diagnostics for the warmer molecular gas. [C{\sc i}](1-0) is also found to be an  excellent alternative H$_2$-gas tracer (\citealt{Papa04,Offn14,Lo14,Bisb15} and \citetalias{Bisb21}). 

The spectral line energy distribution of CO is a key diagnostic for the overall ISM properties \citep{Papa10b,Mash15,Rose15,Vale20}, including its dynamical state \citep{Nara14}. Heating processes such as cosmic-rays, X-rays and turbulence can provide strong volumetric heating and thus increase the gas temperature at column densities where the FUV radiation may be severely attenuated. This results in elevated SLEDs \citep{Brad03,Papa12,Pens21,Espo22}, thus making CO ratios of mid-$J$/(1-0) and above of particular importance. When combined with the total FIR emission, CO SLEDs can further be used to constrain the contribution of photoelectric heating and thus quantify the above volumetric heating processes \citep{Harr21}. In this paper, we examine CO SLEDs up to $J=5-4$.

Finally, it is interesting to explore the [C{\sc ii}]/CO(1-0) and the [C{\sc ii}]/[C{\sc i}](1-0) line ratios. In regards to the first one, earlier observations by \citet{Stac91,Stac10,Hall10} have found that the [C{\sc ii}]/CO(1-0) ratio is approximately $1-4\times10^3$ times higher in starburst than in quiescent galaxies. More recent work by \citet{Accu17b,Madd20} have used that line ratio to infer the CO-dark gas and thus provide a higher estimate of the molecular mass content in galaxies. As we will see below, both [C{\sc ii}]/CO(1-0) and [C{\sc ii}]/[C{\sc i}](1-0) ratios show a similar behaviour for the ISM environmental parameters considered in this work, with [C{\sc ii}]/[C{\sc i}](1-0) having a stronger correlation with the FUV intensity than with the cosmic-ray ionization rate \citep[see also][for the corresponding ratio with {[}C{\sc i}{]}(2-1)]{Pens21}.

\begin{figure}
    \centering
	\includegraphics[width=0.48\textwidth]{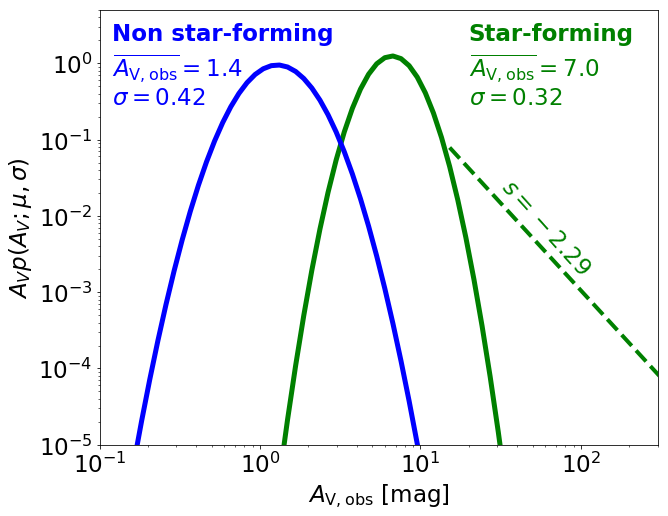}
    \caption{The two $A_{\rm V,obs}$-PDFs considered here. The left one (blue solid line) represents a quiescent, non-star-forming distribution. It has a log-normal (LN) distribution with $\overline{A_{\rm V,obs}}=1.4\,{\rm mag}$ and $\sigma=0.42$. The right one (green lines) represents a star-forming distribution. It has a log-normal distribution with $\overline{A_{\rm V,obs}}=7.0\,{\rm mag}$ and $\sigma=0.32$. For $A_{\rm V,obs}>14\,{\rm mag}$ it has a power-law tail (PL) with slope $s=-2.29$.}
    \label{fig:pdfs}
\end{figure}

\section{Applications}
\label{sec:applications}

Our approach is applied to two different sets of $A_{\rm V,obs}$ distributions. These PDFs are illustrated in Fig.~\ref{fig:pdfs}. The first one (blue colour) corresponds to a quiescent ``non-star-forming" column density distribution. It has a simple log-normal (LN) shape peaking at $\overline{A_{\rm V,obs}}=1.4\,{\rm mag}$ with $\sigma=0.42$. The second one (green colour) corresponds to a ``star-forming" column density distribution. It contains both an LN shape peaking at $\overline{A_{\rm V,obs}}=7.0\,{\rm mag}$ with $\sigma=0.32$, as well as a power-law tail (PL; green dashed line) with slope $s=-2.29$ for $A_{\rm V}\gtrsim14\,{\rm mag}$ and up to ${\sim}500\,{\rm mag}$. In a recent study analysing 120 Milky Way clouds, \citet{Ma22} identified that ${\sim}72\%$ of their sample are LN distributions and ${\sim}18\%$ LN+PL distributions, while the rest ${\sim}10\%$ has an unclear shape. Such PDFs as the aforementioned ones have been frequently observed locally \citep[see][for a such a collection in the Milky Way]{Spil21}. For example, the selected $A_{\rm V,obs}$-PDF of the non-star-forming region is reminiscent to Lupus-V \citep{Kain09} and the one of the star-forming PDF to Cygnus-X \citep{Schn16}. In the extragalactic context, PAWS\footnote{Plateau de Bure Arcsecond Whirlpool Survey.} observations \citep{Hugh13} of the CO $J=1-0$ line in the inner $\sim11\times7\,{\rm kpc}$ region of M51 found\footnote{\citet{Hugh13} convert the CO observations to a $\Sigma_{\rm H2}$ using the standard $X_{\rm CO}=2\times10^{20}\,{\rm cm}^{-2}\,{\rm K}^{-1}\,{\rm km}^{-1}\,{\rm s}$ conversion factor and a 1.36 helium contribution. We convert $\Sigma_{\rm H2}$ to an $A_{\rm V,obs}$ using the $6.3\times10^{-22}\,{\rm mag}\,{\rm cm}^2$ constant and the aforementioned helium contribution.} a lognormal distribution with $\overline{A_{\rm V,obs}}\sim3.25\,{\rm mag}$ and $\sigma=0.44$. This extragalactic $A_{\rm V,obs}$-PDF lies in-between the above `non-star-forming' and `star-forming' distributions.

\subsection{Non star-forming ISM distribution}
\label{ssec:nsf}

\subsubsection{Abundances, gas temperatures and line ratios}
\label{sssec:analysis1}

\begin{figure*}
    \centering
	\includegraphics[width=0.98\textwidth]{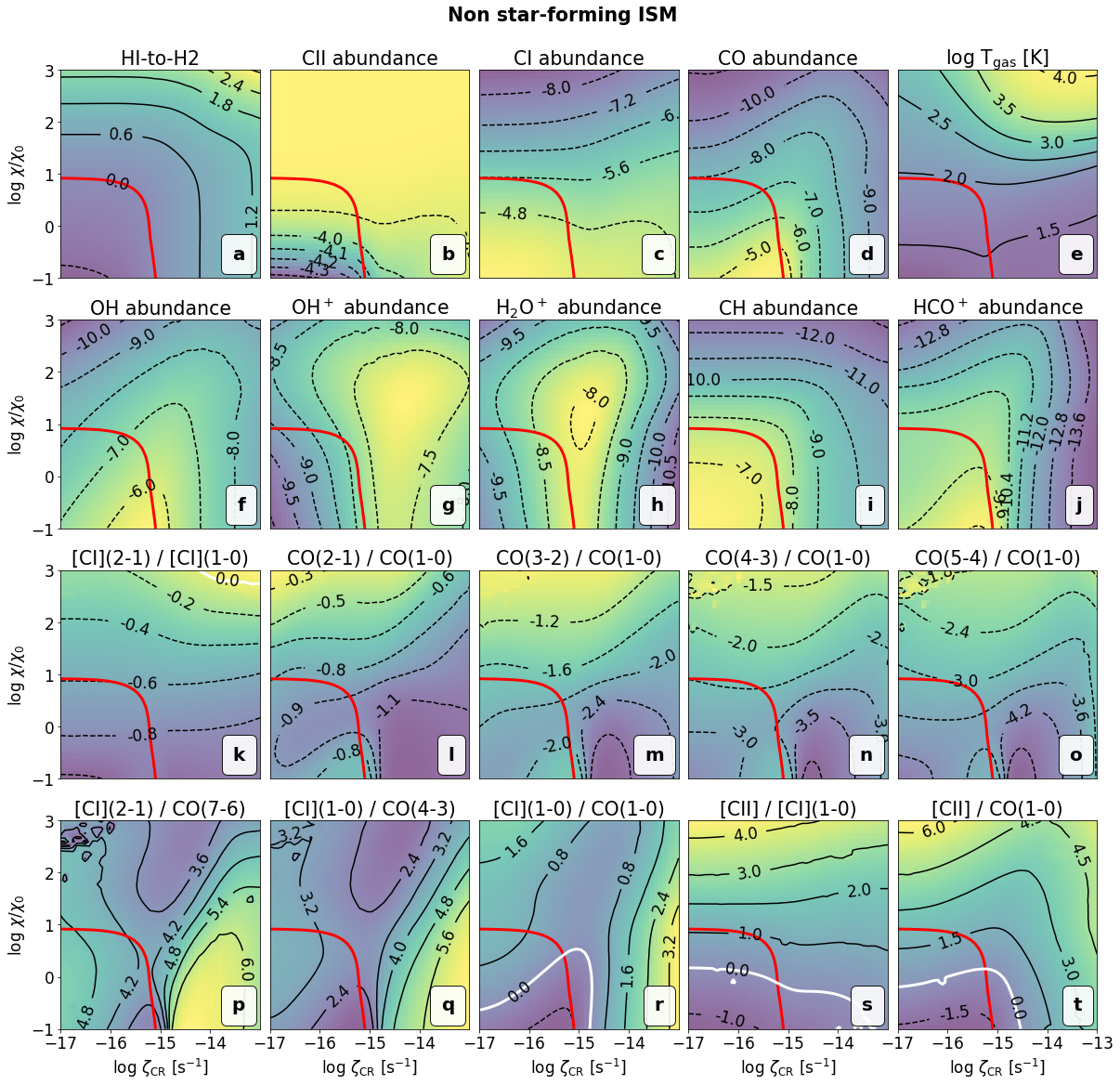}
    \caption{Results for the non-star-forming distribution at $Z=1\,{\rm Z}_{\odot}$. In all panels, $x$-axis is the cosmic-ray ionization rate in the range of $\zeta_{\rm CR}=10^{-17}-10^{-13}\,{\rm s}^{-1}$, and $y$-axis is the FUV intensity in the range of $\chi/\chi_0=10^{-1}-10^3$. All axes and contours shown are in common logarithmic space ($\log_{10}$). Top row from left-to-right: the H{\sc i}-to-H$_2$ transition where the condition H{\sc i}=2H$_2$ is marked in red solid line. This line is overplotted in all panels. The gas is molecular for $\zeta_{\rm CR}$ and $\chi/\chi_0$ pairs smaller than those corresponding to the red line while it is atomic otherwise. The abundances of C{\sc ii}, C{\sc i} and CO are shown in panels~(b-d). Panel~(e) shows the response of the average (density-weighted) gas temperature. The second row (left-to-right) shows the abundances of OH, OH$^+$, H$_2$O$^+$, CH and HCO$^+$. The last two rows show brightness temperature ratios of various emission lines. Whenever shown, the white solid line marks the condition when the brightness temperatures of the two lines in the ratio are equal. Panel~(k) shows the line ratio of [C{\sc i}](2-1)/[C{\sc i}](1-0). The rest of panels in this row illustrate a two-dimensional SLED normalized to CO $J=1-0$, for $J=2-1$ to $5-4$. Bottom row from left-to-right: ratios of [C{\sc i}](2-1)/CO(7-6), [C{\sc i}](1-0)/CO(4-3), [C{\sc i}](1-0)/CO(1-0), [C{\sc ii}]/[C{\sc i}](1-0) and [C{\sc ii}]/CO(1-0).
    }
    \label{fig:lrt}
\end{figure*}

Figure~\ref{fig:lrt} shows the results for the non star-forming ISM distribution for the case of $Z=1\,{\rm Z}_{\odot}$ in metallicity. The x-axis of all panels plots the $\log_{10}\zeta_{\rm CR}$ value, the y-axis the $\log_{10}\chi/\chi_0$ strength of the FUV intensity and all contours are in the $\log_{10}$ space.

Panel~(a) plots the H{\sc i}/2H$_2$ abundance ratio and the red solid line corresponds to the H{\sc i}-to-H$_2$ transition at which the former ratio is equal to one. The distribution is molecular for any $\zeta_{\rm CR}$ and $\chi/\chi_0$ pair that lies underneath the red solid line. As can be seen, this $A_{\rm V,obs}$-PDF remains molecular for a range of cosmic-rays covering up to moderate ionization rates ($\zeta_{\rm CR}\lesssim10^{-15}\,{\rm s}^{-1}$) and FUV intensities ($\chi/\chi_0\lesssim10$). For all other combinations of $\zeta_{\rm CR}$ and $\chi/\chi_0$ the gas becomes atomic. It is interesting to note that when $\zeta_{\rm CR}>10^{-15}\,{\rm s}^{-1}$, H$_2$ is destroyed at approximately equal amounts due to the reactions with H$_2$O$^+$ and OH$^+$ (produced by cosmic-rays) as well as by the direct reaction with cosmic-rays producing H$_2^+$. The aforementioned reactions dominate even in the absence of FUV radiation and thus the medium at such $\zeta_{\rm CR}$ is always atomic. The red solid line is overplotted in all panels of Fig.~\ref{fig:lrt} to guide the eye as to when the medium becomes molecular.

Panels~(b-d) show the abundances of the carbon cycle (C{\sc ii}, C{\sc i} and CO, respectively). For this distribution, the contours of C{\sc ii} and C{\sc i} appear to be approximately horizontal under all $\zeta_{\rm CR}$ values, indicating that they depend primarily on the intensity of FUV radiation. Similarly the abundance of CO, although lower than the abundances of C{\sc ii} and C{\sc i} at all times, show a dependence on the FUV radiation in conditions appropriate for the existence of molecular gas (underneath the red solid line). In the atomic part, the abundance of CO is -as expected- very low but does show a dependence on $\zeta_{\rm CR}$ as we will discuss in more detail in the star-forming ISM case (\S\ref{ssec:sf}).

Panel~(e) shows the density-weighted gas temperature. In the molecular region it varies from $20\lesssim T_{\rm gas}\lesssim100\,{\rm K}$ whereas otherwise it can reach values as high as ${\sim}10^3\,{\rm K}$ for $\chi/\chi_0>10^2$. The average gas temperature, here, depends primarily on the strength of the FUV radiation rather than on the value of $\zeta_{\rm CR}$.

Panel~(f) illustrates the abundance of OH. For conditions favouring the existence of molecular gas, the abundance of OH is $10^{-8}-10^{-6}$ depending on the $\zeta_{\rm CR}$-FUV intensity combination. In particular, OH peaks for $\chi/\chi_0\lesssim1$ and for $10^{-16}\lesssim\zeta_{\rm CR}\lesssim10^{-15}\,{\rm s}^{-1}$. Here, OH forms mainly through dissociative recombination of H$_3$O$^+$. In panels~(g) and (h), the abundances of OH$^+$ and H$_2$O$^+$ are illustrated. Their response in varying $\zeta_{\rm CR}$ and $\chi/\chi_0$ is very similar. Under molecular conditions, they both increase only as a function of $\zeta_{\rm CR}$. Their peak is found to occur for atomic conditions and in particular for $\chi/\chi_0{\sim}30$ and $\zeta_{\rm CR}{\sim}3\times10^{-15}\,{\rm s}^{-1}$. For these values, OH$^+$ forms via the reaction of H$_2$ with O$^+$, and H$_2$O$^+$ forms via the reaction of H$_2$ with OH$^+$. 

Panel~(i) shows contours for the CH abundance. The abundance of this species increases for low $\zeta_{\rm CR}$ and/or $\chi/\chi_0$ peaking for molecular conditions and with both cosmic-rays and FUV intensities minimized. Under these conditions CH is found to be the result both of CH$_2$ reacting with H and through dissociative recombination of CH$_3^+$. In panel~(j), the abundance of HCO$^+$ is plotted. The dependence of this species on $\zeta_{\rm CR}$ and $\chi/\chi_0$ is reminiscent to the CO abundance described above. HCO$^+$ peaks for moderate values of $\zeta_{\rm CR}$ and low FUV intensities and is formed via CO$^+$ reacting with H$_2$. 

The bottom two rows show the results for the line ratios. In particular in panel~(k), the atomic carbon line ratio ([C{\sc i}](2-1)/[C{\sc i}](1-0)) is plotted. Under all combinations of $\zeta_{\rm CR}$ and $\chi/\chi_0$ explored, this line ratio depends strongly on the FUV intensity and is nearly independent of the cosmic-ray ionization rate. It can thus be used as a tracer for the value of $\chi/\chi_0$ for a column density distribution similar to the non-star-forming case. It appears that when the two environmental parameters explored are maximized, the antenna temperature of [C{\sc i}](2-1) becomes stronger than [C{\sc i}](1-0). However, for such extreme conditions the gas is heavily atomic with the carbon to be mainly in the ionized phase; thus the emission of both [C{\sc i}] lines may not be detectable.

\begin{figure*}
    \centering
	\includegraphics[width=0.98\textwidth]{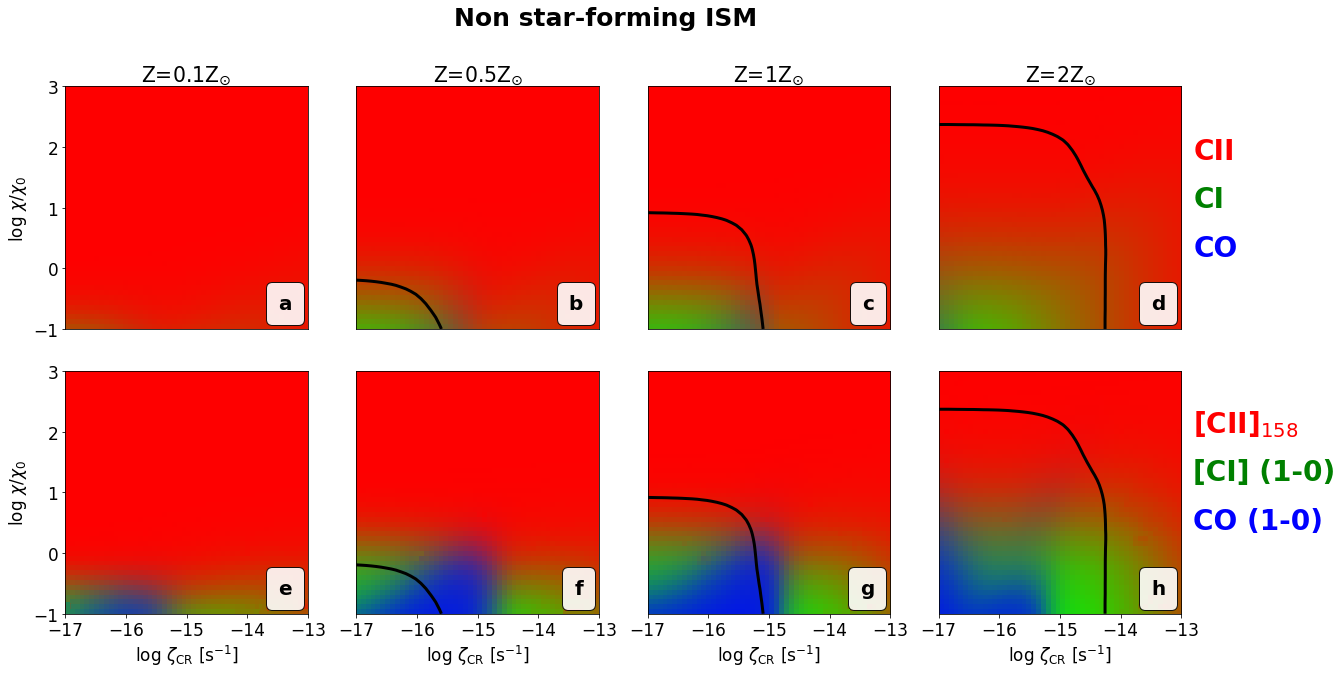}
    \caption{Maps showing which carbon phase dominates for all combinations of the ISM environmental parameters considered. The $x$-axis is the cosmic-ray ionization rate and the $y$-axis is the FUV intensity. These maps are calculated for the non-star-forming $A_{\rm V,obs}$-PDF. The top row shows which carbon phase dominates in terms of the abundance while the bottom row shows the same but in terms of the brightness temperature. The solid black line shows the H{\sc i}-to-H$_2$ transition, below which the distribution is molecular, otherwise it becomes atomic. Red colour is for C{\sc ii} and the [C{\sc ii}]~$158\mu$m line, green colour for the C{\sc i} abundance and the [C{\sc i}](1-0) $609\mu$m line, and blue colour for the CO abundance and the CO(1-0) line. Each column corresponds to different metallicity ($Z=$0.1, 0.5, 1 and 2 Z$_{\odot}$, from left-to-right). As can be seen, the fact that a particular carbon phase dominates in terms of the abundance, does not always mean that it will also have the brightest emission, since gas temperature and optical depth effects need to be accounted for.}
    \label{fig:lf}
\end{figure*}

In panels~(l-o), a two-dimensional CO SLED is illustrated. Focusing on the ratios for molecular conditions, it can be seen that all $(J\rightarrow J-1)/(1-0)$ ratios obtain their highest value for low FUV intensities ($\chi/\chi_0\lesssim1$) and moderate-to-low cosmic-ray ionization rates ($10^{-16}\lesssim\zeta_{\rm CR}\lesssim10^{-15}\,{\rm s}^{-1}$). For these conditions and the particular $A_{\rm V,obs}$-PDF, the CO abundance obtains its highest value while the average gas temperature is ${\sim}30\,{\rm K}$, leading to an elevated antenna temperature of the mid-$J$ lines when compared to the rest of ISM conditions. This trend can be better seen and understood in the star-forming $A_{\rm V,obs}$-PDF case which is explained in more detail in \S\ref{sssec:analysis2}.

Finally, in the fourth row, five of the most frequently used ratios are shown. As before, the focus here will be for the conditions leading to molecular gas. Panel~(p) shows the [C{\sc i}](2-1)/CO(7-6) ratio which indicates a much brighter [C{\sc i}](2-1) antenna temperature than CO(7-6). This ratio is ${\sim}6\times10^4$ for low $\zeta_{\rm CR}$ but decreases to ${\sim}4\times10^3$ for $\zeta_{\rm CR}{\sim}10^{-15}\,{\rm s}^{-1}$, since cosmic-rays can elevate the gas temperature at high column densities increasing the CO(7-6) line emission. Similarly, the [C{\sc i}](1-0)/CO(4-3) ratio shown in panel~(q), also indicates a ${\sim}10^2-10^3$ times brighter [C{\sc i}](1-0) antenna temperature. Such high ratios are estimated because high-$J$ CO are brightly emitted only from the high-end of $A_{\rm V,obs}$-PDF and are thus overall very weak.

The [C{\sc i}](1-0)/CO(1-0) ratio shown in panel~(r) is of particular interest. The white solid line marks the case when the antenna temperatures of these two emission lines are the same. The dependence of this ratio on the $\zeta_{\rm CR}$ and $\chi/\chi_0$ parameters is reminiscent to that of the CO abundance. In particular, it is found that the antenna temperature of CO(1-0) is brighter than that of [C{\sc i}](1-0) for generally low FUV intensities and for $\zeta_{\rm CR}{\sim}2\times10^{-15}\,{\rm s}^{-1}$, which surpasses the H{\sc i}-to-H$_2$ transition as can be seen (see Appendix~\ref{sssec:supra}). The [C{\sc ii}]/[C{\sc i}](1-0) ratio illustrated in panel~(s) shows that this non-star-forming distribution is in general bright in [C{\sc i}](1-0) and becomes brighter in [C{\sc ii}] when $\chi/\chi_0$ increases. This increase in [C{\sc ii}] is expected, since the FUV radiation ionizes carbon to form C{\sc ii} and also increases the gas temperature. Finally, the [C{\sc ii}]/CO(1-0) ratio pattern shown in panel~(t) is also reminiscent to the one of CO abundance. As with the [C{\sc ii}]/[C{\sc i}](1-0) ratio, the [C{\sc ii}] antenna temperature is higher than the CO(1-0) one as $\chi/\chi_0$ increases.  Interestingly, all three ratios explored between CO(1-0), [C{\sc i}](1-0) and [C{\sc ii}], have their ratio equal to 1 for $\chi/\chi_0{\sim}1$ regardless to the $\zeta_{\rm CR}$ value.

Appendix~\ref{sssec:tau_nSF} discusses the behaviour of the optical depths of [C{\sc ii}], [C{\sc i}](1-0) and CO(1-0) for the non-star-forming ISM distribution. Figures~\ref{fig:lrt0p1}-\ref{fig:lrt2p0} of Appendix~\ref{app:abnd} show the corresponding results of abundances and line ratios for $Z=0.1$, 0.5 and $2.0\,{\rm Z}_{\odot}$, respectively.

\subsubsection{Carbon phases at different metallicities}
\label{sssec:cphase1}

Figure~\ref{fig:lf} shows which carbon  phase dominates for all the ISM environmental parameters considered in this work. The top row shows the response of the abundances of C{\sc ii} (red), C{\sc i} (green) and CO (blue) while the bottom row shows their corresponding antenna temperatures ([C{\sc ii}]~158$\mu$m, [C{\sc i}](1-0), CO(1-0), respectively). In all panels, the black line shows the H{\sc i}-to-H$_2$ transition. Each column corresponds to different metallicities, from $Z=0.1\,{\rm Z}_{\odot}$ to $2\,{\rm Z}_{\odot}$ (left-to-right). 

In regards to the abundances (top row of Fig.~\ref{fig:lf}), for $Z=0.1\,{\rm Z}_{\odot}$ the gas is purely atomic under all ISM conditions as shown in panel~(a). Carbon is mainly found in C{\sc ii} form, since its abundance dominates for all $\zeta_{\rm CR}$-FUV intensity combinations. The non star-forming distribution appears to become molecular only for $Z>0.5\,{\rm Z}_{\odot}$ (panel~b) and for a combination of very low amounts of FUV radiation ($\chi/\chi_0\lesssim1$) and low $\zeta_{\rm CR}$ ($\lesssim10^{-16}\,{\rm s}^{-1}$). As metallicity increases to solar and super-solar values (panels~c-d), the distribution remains molecular for combinations of higher $\chi/\chi_0$ and $\zeta_{\rm CR}$. This is because FUV radiation is more strongly attenuated due to the increased dust shielding allowing H$_2$ molecules to form more efficiently at lower column densities. For all metallicity cases, the distribution never becomes CO-dominated but rather C{\sc i}-dominated and only for low $\chi/\chi_0$ and $\zeta_{\rm CR}$. Otherwise it is always C{\sc ii}-dominated.

The maps of carbon cycle abundances are not always reflected in the corresponding line emission maps as can be seen in the bottom row of Fig.~\ref{fig:lf}. For $Z=0.1\,{\rm Z}_{\odot}$ (panel~e) it is found that for $\chi/\chi_0\lesssim0.3$ and $\zeta_{\rm CR}\lesssim2\times10^{-16}\,{\rm s}^{-1}$, the distribution, although atomic, can be bright in CO(1-0) and also [C{\sc i}](1-0). In particular, as can be seen also for $Z=0.5$ and $1\,{\rm Z}_{\odot}$ (panels~f-g), the atomic gas can be indeed bright in the aforementioned lines. There are two reasons that explain this feature: i) the $A_{\rm V,obs}$-PDF contains high column densities and although they have small probability, they can be bright in CO(1-0) and [C{\sc i}](1-0) and ii) the suprathermal formation route of CO via CH$^+$ considered in these runs can increase the abundance of CO at low visual extinctions and thus the corresponding level populations, building a bright CO(1-0) antenna temperature at low $A_{\rm V}$. The latter effect is further discussed in Appendix~\ref{sssec:supra}. 

It is further found that the molecular distribution can be [C{\sc i}](1-0) as well as [C{\sc ii}]~$158\mu$m bright (panels~g-h). In general, it will become [C{\sc ii}] bright as the FUV intensity increases -and, in this $A_{\rm V,obs}$-PDF case- only when metallicities $Z\gtrsim1\,{\rm Z}_{\odot}$. In addition, as the $\chi/\chi_0$ increases, the CO(1-0) brightness temperature will be followed by [C{\sc i}](1-0) and then to [C{\sc ii}]~$158\mu$m, although for moderate $\zeta_{\rm CR}$ values the [C{\sc i}](1-0) brightness temperature never dominates (the gas is either CO(1-0) or [C{\sc ii}] bright). For low $\chi/\chi_0$ and as $\zeta_{\rm CR}$ increases, the aforementioned transition sequence of CO(1-0) to [C{\sc i}](1-0) to [C{\sc ii}] is also repeated due to the CO destruction. 

In panels~\ref{fig:app_flags}a,c of Appendix~\ref{app:avfac}, it is shown how the above results for $Z=1\,{\rm Z}_{\odot}$ would differ if a lower $A_{\rm V}/N_{\rm H}$ factor was adopted.

\begin{figure*}
    \centering
	\includegraphics[width=0.98\textwidth]{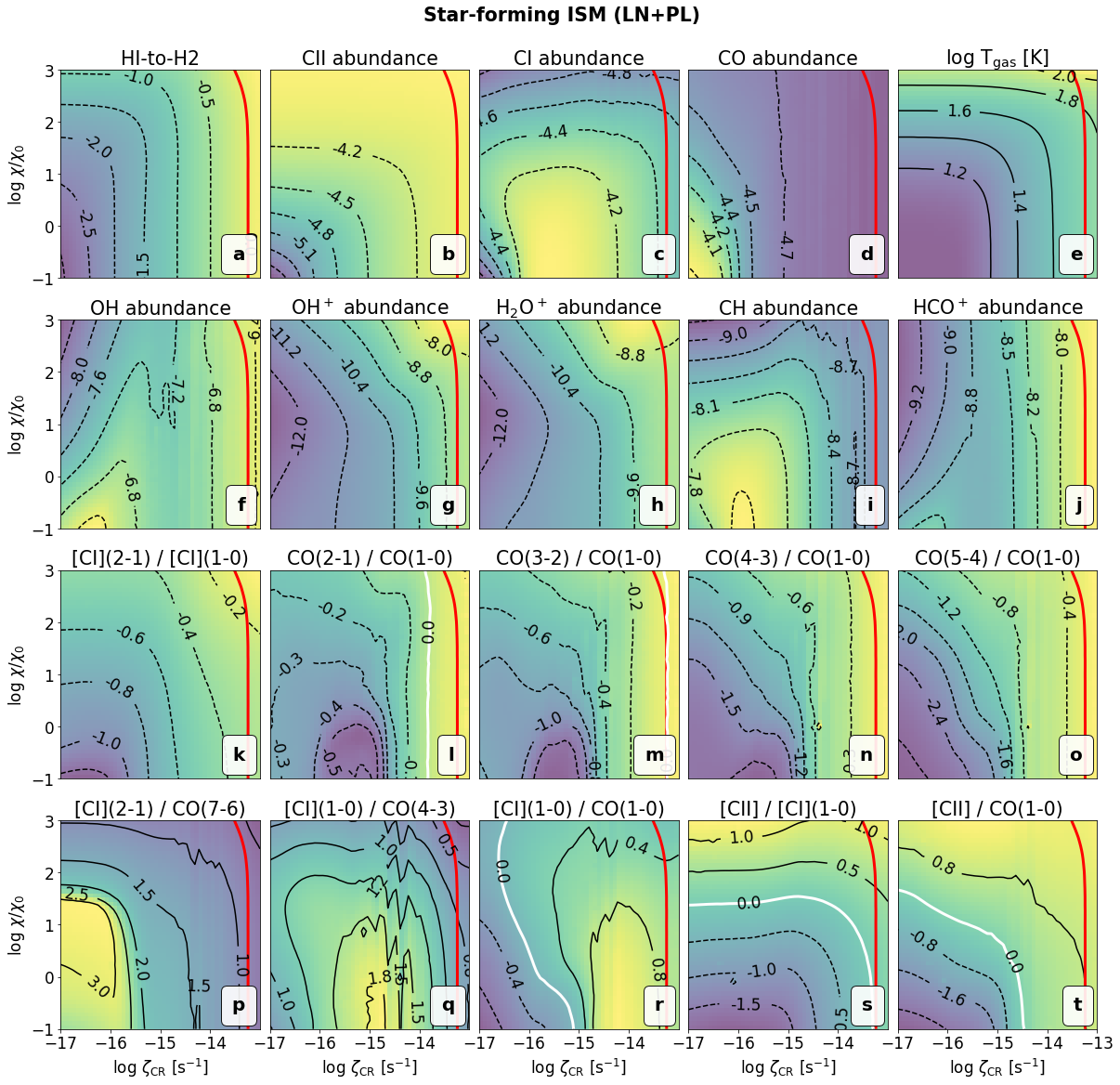}
    \caption{As in Fig.~\ref{fig:lrt} but for the $A_{\rm V,obs}$-PDF of the star-forming ISM at solar metallicity, including both LN and PL components.}
    \label{fig:crt}
\end{figure*}

\subsection{Star-forming ISM distribution}
\label{ssec:sf}

\subsubsection{Abundances, gas temperatures and line ratios}
\label{sssec:analysis2}

Following the description of Fig.~\ref{fig:lrt} for the non-star-forming case, Fig.~\ref{fig:crt} shows the results for the star-forming ISM distribution at $Z=1\,{\rm Z}_{\odot}$. Here, both the lognormal (LN) and power-law (PL) components have been considered. As can be seen in panel~(a), this distribution remains largely molecular for all combinations of $\zeta_{\rm CR}$ and FUV intensities explored, except for very high cosmic-rays i.e. $\zeta_{\rm CR}\gtrsim5\times10^{-14}\,{\rm s}^{-1}$. 

The carbon cycle illustrated in panels~(b-d), has an interesting behaviour depending on the $\zeta_{\rm CR}$-FUV intensity combination. The C{\sc ii} abundance increases both when the FUV intensity increases as a result of atomic carbon photoionization and when the cosmic-ray ionization rate increases as a result of the destruction of CO by He$^+$, peaking for high values of $\zeta_{\rm CR}$ and/or $\chi/\chi_0$. For $\chi/\chi_0>10$, the abundance of C{\sc i} decreases and depends almost entirely on the value of FUV intensity. However, for $\chi/\chi_0<10$, it increases and peaks for moderate cosmic-rays e.g. $10^{-16}<\zeta_{\rm CR}<3\times10^{-15}\,{\rm s}^{-1}$, as a result of the cosmic-ray induced destruction of CO \citep{Bisb15,Bisb17b}. In this effect cosmic-rays create a surplus of He$^+$ which reacts very effectively with CO creating C{\sc ii}. In moderate $\zeta_{\rm CR}$, C{\sc ii} recombines quickly creating large amounts of C{\sc i}; hence the C{\sc i} abundance peak seen in the corresponding panel. At high $\zeta_{\rm CR}$, this recombination is not so effective and thus carbon remains in the form of C{\sc ii}. The aforementioned sequence is also reflected in the CO abundance panel which shows a strong dependence on $\zeta_{\rm CR}$ rather than on $\chi/\chi_0$. During all the above carbon cycle phase changes, the gas remains always molecular.

The average, density-weighted gas temperature illustrated in panel~(e), shows a dependence on both $\zeta_{\rm CR}$ and $\chi/\chi_0$. Like in the non star-forming distribution, the gas temperature is found to be in the range of $10\lesssim T_{\rm gas}\lesssim60\,{\rm K}$. For a constant $\zeta_{\rm CR}$, the gas temperature increases as a result of the increase of photoelectric heating. For a constant $\chi/\chi_0$, the gas temperature increases as result of the increase of cosmic-ray, chemical and H$_2$ formation heating mechanisms \citep{Bisb17b}. The particular chemical heating is a result of contributions by photoelectrons, turbulence dissipation, exothermic reactions due to HCO$^+$, H$_3^+$ and H$_3$O$^+$ recombination as well as ion-neutral reactions between H$_2^+$ and He$^+$.

Panel~(f) shows the response of OH abundance. For low $\zeta_{\rm CR}$ (i.e. $\lesssim10^{-16}\,{\rm s}^{-1}$) and high FUV intensities ($\chi/\chi_0>10$), the abundance of OH is reduced, whereas for all other cases and especially for high $\zeta_{\rm CR}$ or when $\chi/\chi_0$ is very low, it is increased, as a result of the H$_3$O$^+$ dissociative recombination. Furthermore, as can be seen, the abundance of OH strongly depends on $\zeta_{\rm CR}$ particularly when $\zeta_{\rm CR}\gtrsim10^{-15}\,{\rm s}^{-1}$. Similarly, both OH$^+$ and H$_2$O$^+$ species (panels~g,h) depend strongly on cosmic-rays and weakly on $\chi/\chi_0$. The abundance of the latter two ions is high for a combination of high $\zeta_{\rm CR}$ and high $\chi/\chi_0$. On the other hand the abundance of CH, as seen in panel~(i), peaks for low FUV intensities ($\chi/\chi_0<1$) and for $\zeta_{\rm CR}\simeq10^{-16}\,{\rm s}^{-1}$ as a result of the CH$_2$~+~H reaction and CH$_3^+$ recombination. It is reduced when $\chi/\chi_0$ and/or $\zeta_{\rm CR}$ increase. In panel~(j), the abundance of HCO$^+$ is illustrated which shows a strong dependence on the value of $\zeta_{\rm CR}$ while having a negligible dependence on the value of FUV intensity. 

The [C{\sc i}](2-1)/[C{\sc i}](1-0) line ratio illustrated in panel~(k), indicates that in systems with column density distributions reminiscent to the star-forming ISM modelled here, it may be used to constrain the value of $\zeta_{\rm CR}$. As can be seen, this line ratio depends on $\zeta_{\rm CR}$ particularly when $\zeta_{\rm CR}\gtrsim10^{-15}\,{\rm s}^{-1}$, a value that may be met in starburst galaxies and systems with high star-formation rates \citep{Bisb21}. For cosmic-rays with $\zeta_{\rm CR}\lesssim10^{-15}\,{\rm s}^{-1}$, this ratio depends almost entirely on $\chi/\chi_0$ and thus can be used to constrain the FUV intensity. 

The two-dimensional CO SLED shown in panels~(l-o), indicate that the $(J\rightarrow J-1)/(1-0)$ CO ratio shows a strong dependence on cosmic-rays for $\zeta_{\rm CR}\gtrsim10^{-15}\,{\rm s}^{-1}$. As described above, this is due to the fact that high cosmic-ray energy densities heat up the gas at high column densities, thus making mid-$J$ and high-$J$ CO transitions very bright (see \S\ref{ssec:ratios}). Note that the particular CO(2-1)/CO(1-0) is $>1$ for $\zeta_{\rm CR}\gtrsim10^{-14}\,{\rm s}^{-1}$. Such a case is also observed in the CO(3-2)/CO(1-0) ratio for even higher $\zeta_{\rm CR}$, at which point however the distribution may become more atomic-dominated. 

\begin{figure*}
    \centering
	\includegraphics[width=0.98\textwidth]{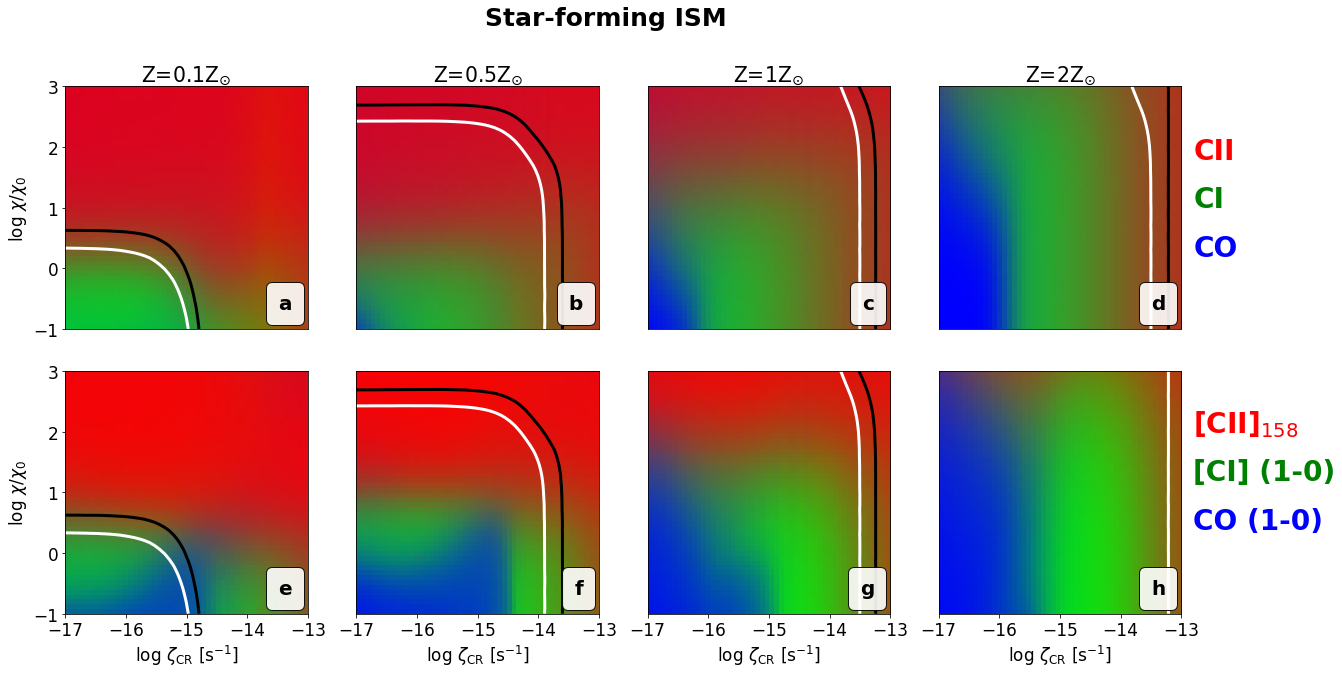}
    \caption{As in Fig.~\ref{fig:lf} for the star-forming ISM. The coloured results and the black solid line of the H{\sc i}-to-H$_2$ transition are for the LN+PL case. The white solid line corresponds to the results for the LN only case. For the carbon cycle abundances and emission, there are no appreciable differences between the LN+PL and LN cases.}
    %White lines denote the corresponding results for the LN only case. In particular, the white solid line is the H{\sc i}-to-H$_2$ transition, the white dashed line is the boundary of the C{\sc i} species and the white dotted line is the boundary of the CO species.}
    \label{fig:cf}
\end{figure*}

In the bottom row, the [C{\sc i}](2-1)/CO(7-6) ratio is illustrated in panel~(p). As with the non-star-forming ISM distribution (\S\ref{sssec:analysis1}), [C{\sc i}](2-1) is brighter than CO(7-6). In particular [C{\sc i}](2-1) is $\sim500-1000$ times brighter for $\chi/\chi_0\lesssim50$ and $\zeta_{\rm CR}\lesssim2\times10^{-16}\,{\rm s}^{-1}$ whereas it is just a few tens brighter otherwise. The [C{\sc i}](1-0)/CO(4-3) ratio shown in panel~(q)\footnote{Note on the instabilities of panel~(q) for $\zeta_{\rm CR}\sim10^{-15}-10^{-14}\,{\rm s}^{-1}$: numerical instabilities arise from the non-linear nature of the ordinary differential equations that are solved to achieve equilibrium in PDRs. Early work of \citet{LeBo93} demonstrated the existence of bi-stability in such solutions in dark molecular gas. Subsequent papers \citep[e.g.][]{Viti01, Roue20, Dufo21} have also addressed and examined further bi-stability, especially in the oxygen chemistry. The instabilities seen in panel~(q) (and elsewhere) is the result of the above.}, peaks for moderate $\zeta_{\rm CR}$ ($\sim10^{-15}\,{\rm s}^{-1}$) for low FUV intensities with [C{\sc i}](1-0) to be $\sim30$ times brighter than CO(4-3). Both aforementioned line ratios show a stronger correlation with $\zeta_{\rm CR}$ than with $\chi/\chi_0$. 

Finally, in panels~(r-t), the [C{\sc i}](1-0)/CO(1-0), [C{\sc ii}]/[C{\sc i}](1-0) and [C{\sc ii}]/CO(1-0) are shown, respectively. The first ratio shows a stronger correlation with $\zeta_{\rm CR}$ than the other two which are more sensitive on the FUV intensity. The antenna temperature of [C{\sc i}](1-0) is approximately equal to the CO(1-0) for $3\times10^{-17}\lesssim\zeta_{\rm CR}\lesssim10^{-15}\,{\rm s}^{-1}$ depending on $\chi/\chi_0$; the lower the FUV intensity, the higher the cosmic-ray ionization rate needs to be to satisfy the equality of the [C{\sc i}](1-0) and CO(1-0) antenna temperatures. For smaller $\zeta_{\rm CR}$, CO(1-0) becomes brighter. Evidently, in systems containing high column densities and high cosmic-ray energy densities, [C{\sc i}](1-0) can be $4-6$ times brighter than CO(1-0) as indicated in the middle panel. In regards to the [C{\sc ii}]/[C{\sc i}](1-0) ratio, it appears to be more sensitive to the $\chi/\chi_0$ value than the $\zeta_{\rm CR}$ one. For $\chi/\chi_0\lesssim20$, this star-forming ISM distribution is brighter in [C{\sc i}](1-0) than [C{\sc ii}]. Similarly, as can be seen in the last panel, CO(1-0) is brighter than [C{\sc ii}] for any combination satisfying $\chi/\chi_0\lesssim10$ and $\zeta_{\rm CR}\lesssim2\times10^{-15}\,{\rm s}^{-1}$. In all other cases, the molecular gas may be primarily [C{\sc ii}]-bright. 

Appendix~\ref{sssec:tau_SF} discusses the behaviour of the optical depths for the star-forming ISM distribution. Figures~\ref{fig:crt0p1}-\ref{fig:crt2p0} show the corresponding results for $Z=0.1$, 0.5 and $2.0\,{\rm Z}_{\odot}$, respectively.

\subsubsection{Carbon phases at different metallicities}
\label{sssec:cphase2}

Figure~\ref{fig:cf} shows which carbon phase dominates for all ISM environmental parameters explored. The coloured results along with the black solid line of the H{\sc i}-to-H$_2$ transition correspond to the case where the power-law tail is included (LN+PL). The white solid line corresponds to the results where only the log-normal component is included (see \S\ref{sssec:PL} for the latter). 

In regards to the abundances (top row), the H{\sc i}-to-H$_2$ transition changes with metallicity (see also \S\ref{sssec:cphase1}), however this star-forming distribution remains molecular even at very low ($0.1\,{\rm Z}_{\odot}$) metallicities (panel~a). Here, carbon is mostly found in the form of C{\sc i} in the molecular phase and C{\sc ii} otherwise. For $Z=0.5\,{\rm Z}_{\odot}$ (panel~b) the molecular phase is mostly C{\sc i} dominated for $\chi/\chi_0\lesssim1$ (except for a negligible intensity of FUV radiation and for very low $\zeta_{\rm CR}$ in which it is CO-dominated) but transitions to C{\sc ii} for $\zeta_{\rm CR}\gtrsim10^{-14}\,{\rm s}^{-1}$. For $\chi/\chi_0\gtrsim1$ it is already C{\sc ii}-dominated. At $Z=1\,{\rm Z}_{\odot}$ (panel~c), the C{\sc i}-dominated phase holds for $\chi/\chi_0\simeq10$ but for $\zeta_{\rm CR}\lesssim10^{-16}\,{\rm s}^{-1}$, the majority of carbon is found in CO form. Interestingly, for $\zeta_{\rm CR}\simeq10^{-17}\,{\rm s}^{-1}$, the carbon phase may switch from CO directly to C{\sc ii} as $\chi/\chi_0$ increases, without a C{\sc i}-dominated phase (see also `Case-2' of \citetalias{Bisb19}). Finally, when $Z=2\,{\rm Z}_{\odot}$ (panel~d), the star-forming distribution is found to remain molecular under all ISM conditions considered. The CO-dominated phase can exist for much higher FUV intensities, but it always transitions to C{\sc i} and then to C{\sc ii} as $\zeta_{\rm CR}$ increases to $\gtrsim1-3\times10^{-16}\,{\rm s}^{-1}$ due to the cosmic-ray induced CO destruction effect. Notably, the C{\sc i}-dominated phase exists for a wide range of FUV intensities and metallicities but always transitions to C{\sc ii} when $\zeta_{\rm CR}\gtrsim10^{-14}\,{\rm s}^{-1}$.

In regards to the brightness temperatures of the carbon cycle species (bottom row), for $Z=0.1\,{\rm Z}_{\odot}$ (panel~e) the molecular phase is in general bright in [C{\sc i}](1-0). For very low FUV intensities it is the antenna temperature of CO(1-0) that dominates depending on $\zeta_{\rm CR}$, even though it is rich in C{\sc i} abundance as seen in the top left panel. Interestingly, for $\chi/\chi_0\lesssim1$ the distribution remains [C{\sc i}](1-0) bright even for $\zeta_{\rm CR}$ values for which the gas is in atomic form. At $Z=0.5\,{\rm Z}_{\odot}$ (panel~f), the molecular phase is bright in [C{\sc i}](1-0) for $\chi/\chi_0\lesssim5$, except when $\chi/\chi_0\lesssim1$ and $\zeta_{\rm CR}\lesssim3\times10^{-15}\,{\rm s}^{-1}$ in which case it is bright in CO(1-0). For any stronger FUV intensity, the molecular gas is mainly [C{\sc ii}]-bright. As with the $Z=0.1\,{\rm Z}_{\odot}$ case, the medium is almost nowhere rich in CO abundance, although its line emission in the $J=1-0$ transition can be strong. 

At $Z=1\,{\rm Z}_{\odot}$ (panel~g), the gas is [C{\sc ii}]-bright for $\chi/\chi_0\gtrsim20$, regardless to the $\zeta_{\rm CR}$ value, since the gas is rich in C{\sc ii} abundance due to photodissociation of CO. For lower FUV intensities, the gas is CO(1-0) bright for $\zeta_{\rm CR}\lesssim10^{-16}\,{\rm s}^{-1}$ on average. Note that for $\zeta_{\rm CR}{\sim}10^{-17}\,{\rm s}^{-1}$, the gas transitions from CO(1-0)-bright to [C{\sc ii}]-bright, as well, directly when the FUV intensity exceeds approximately $20\chi_0$. As $\zeta_{\rm CR}$ increases and for this low FUV intensity regime, this ISM distribution remains [C{\sc i}](1-0) even for very high cosmic-rays. Finally, for $Z=2\,{\rm Z}_{\odot}$ (panel~h), the gas is in practice either CO(1-0)-bright (for $\zeta_{\rm CR}\lesssim3\times10^{-16}\,{\rm s}^{-1}$) or [C{\sc i}](1-0)-bright otherwise. It may become [C{\sc ii}] only for very extreme ISM environmental parameters. Note that for the latter two metallicity cases, the abundances (top third and fourth panels) and the line emission (bottom third and fourth panels) patterns agree well with each other.

\subsubsection{Contribution of the power-law tail to the PDR results}
\label{sssec:PL}

As can be seen from Fig.~\ref{fig:cf}, the power-law tail increases the contribution of the high column density material to the total one and, as such, it increases the total molecular gas mass content. This can be seen from the combination of the ISM conditions that are required for the H{\sc i}-to-H$_2$ transition to occur; in the LN case (white solid line) it occurs for somehow lower $\zeta_{\rm CR}$ and lower $\chi/\chi_0$ when compared to the LN+PL case (black solid line). Overall, we find that the PL tail retains molecular conditions for twice the FUV intensity and twice the cosmic-ray ionization rate than the LN-only case. 

We find that both the abundances and the line emissions of the carbon cycle remain in general very similar for the two cases. However, the PL tail slightly increases the CO abundance for higher $\chi/\chi_0$ (not plotted; see also \citetalias{Bisb19}). 

In panels~\ref{fig:app_flags}b,d of Appendix~\ref{app:avfac}, it is shown how the above results for $Z=1\,{\rm Z}_{\odot}$ would differ if a lower $A_{\rm V}/N_{\rm H}$ factor was adopted.

\section{Discussion}
\label{sec:discussion}

{\sc PDFchem} provides a fast calculation of key abundances and emission line ratios that are most commonly used, for large-scale (tens-to-hundreds of pc) inhomogeneous clouds characterized by an $A_{\rm V,obs}$-PDF. As described earlier, while many such distributions have been obtained for regions in the Milky Way, the limited resolution for extragalactic systems does not allow to determine such PDFs at high-redshift e.g. in the crucial $z\sim2-3$ range for the cosmic star-formation theory \citep{Mada14}. In terms of spatial resolution, the ISM at $z\sim2-3$ is observed at a sub-kpc resolution \citep[see][for a review with ALMA]{Hodg20}, unless gravitational lensing effects occur which may reveal scales of $\sim100-200\,{\rm pc}$ \citep[e.g.][]{Ryba20,Soli21}. It would be therefore interesting to use {\sc PDFchem} and input educated guesses of $A_{\rm V,obs}$ distributions to explore the trends we may expect when studying extragalactic objects, particularly high-redshift galaxies. 

Currently, such aforementioned educated guesses can arise from hydrodynamical models. For instance, simulations of the turbulent, star-forming ISM have explored the development of column density PDFs \citep[e.g.][]{Kles00,Fede13,Burk19}. Particular attention has been paid on the transition of lognormal to power-law component, with the models suggesting that the power-law tail is a product of free-fall collapse forming filaments and dense clumps \citep{Krit11,Giri14} including contribution from magnetic fields \citep{Audd18}. Overall, the majority of literature focuses on the description of $A_{\rm V,obs}$-PDFs resulting from GMC-scale models. Having shown in this work that such PDFs are valuable tools for a quick examination of the atomic/molecular content and their corresponding line emissions, we encourage the modelling community to enrich the current literature with more results and analyses on $A_{\rm V,obs}$-PDFs especially from galaxy-sized simulations.

\subsection{Tracing molecular gas under different conditions}

As discussed previously, the lower panels of Figs.~\ref{fig:lf} and \ref{fig:cf} illustrate how the dominant carbon phase emitter --under molecular conditions-- changes as a function of the ISM environmental parameters. The carbon cycle transitions are found to be more strongly correlated with the ISM conditions, than the H{\sc i}-to-H$_2$ transition is. Consequently, the ISM parameters affect the methodology of tracing the H$_2$ gas as well as the corresponding conversion factors. 

\begin{figure*}
    \centering
	\includegraphics[width=0.87\textwidth]{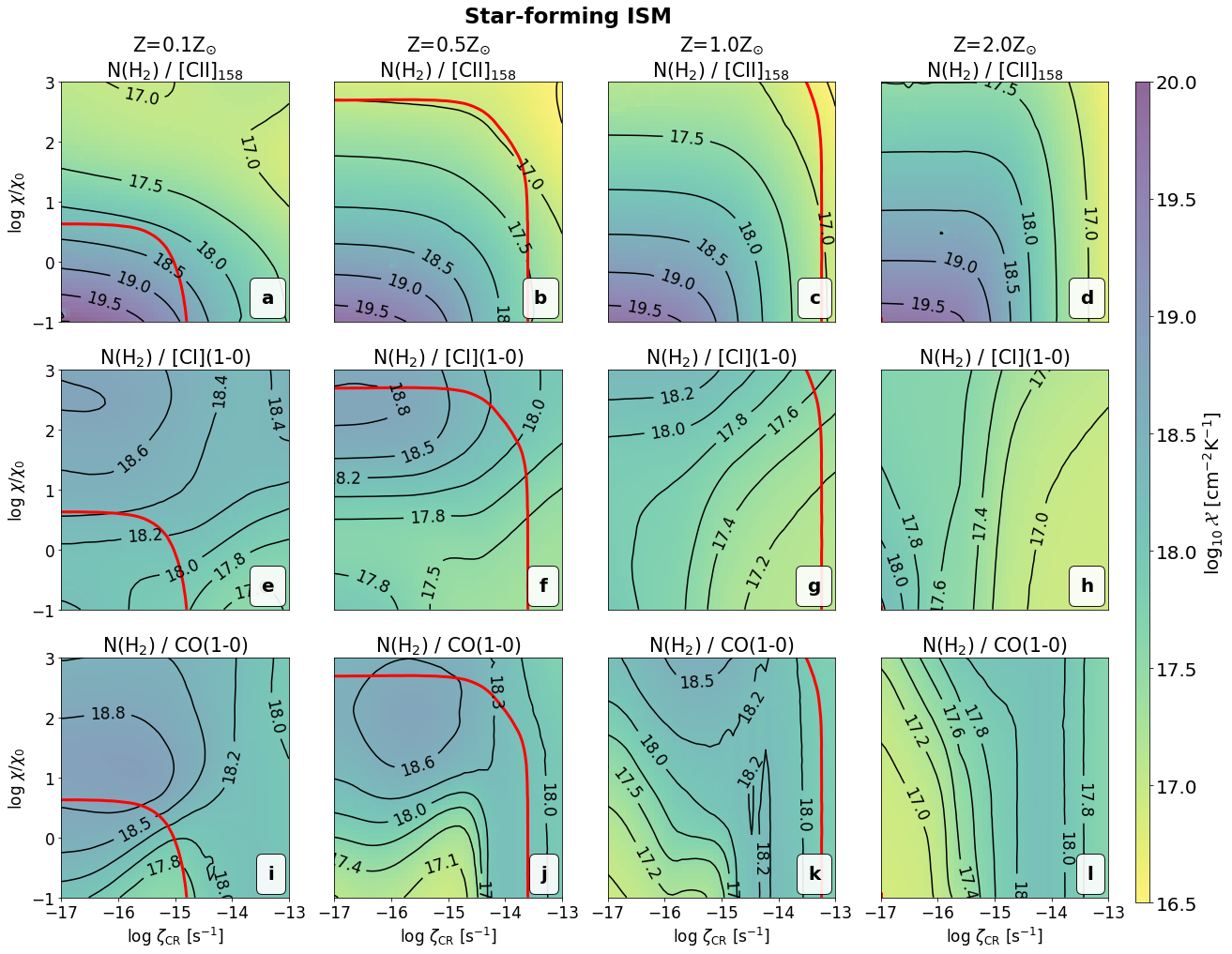}
    \caption{Logarithmic grid maps showing ratios of the PDF-averaged H$_2$ column density to the PDF-averaged antenna temperature of each tracer (units of $\rm cm^{-2}\,K^{-1}$). Each column corresponds to a different metallicity (left-to-right: $Z=0.1$, 0.5, 1.0, $2.0\,{\rm Z}_{\odot}$) and each row to a different tracer (top-to-bottom: [C{\sc i}]~$158\mu$m, [C{\sc i}](1-0), CO $J=1-0$). The star-forming $A_{\rm V,obs}$-PDF has been used in all cases. The colour bar shows the logarithm of the $\cal X$-factor.}
    \label{fig:conversions}
\end{figure*}

In this work, we introduce the quantity ${\cal X}_*\equiv\rm \langle N(H_2)\rangle/\langle T_*\rangle$ representing the ratio of the PDF-averaged H$_2$ column density and the PDF-averaged antenna temperature of the emitter. This ratio has units of [$\rm cm^{-2}\,K^{-1}$] and it differs from the standard `$X$-factor' by the linewidth. $\cal X$ is used here to illustrate the expected behaviour of the commonly used conversion factors (e.g. the CO- or the CI-to-H$_2$) under the different ISM conditions. The reader should be aware that the presented results are further prone to the choice of the inputted column density PDF and they should, thus, only be considered to demonstrate qualitatively (rather than quantitatively) the trends of $X$-factors. Figure~\ref{fig:conversions} shows a collective map of how $\cal X$ changes as a function of the ISM environmental parameter explored. The star-forming $A_{\rm V,obs}$-PDF was adopted for these calculations, to best represent a dense molecular region. The top row of Fig.~\ref{fig:conversions} shows the $\cal X_{\rm [CII]}$ ratio while the middle and bottom rows the $\cal X_{\rm [CI](1-0)}$ and $\cal X_{\rm CO(1-0)}$, respectively. Each column corresponds to different metallicity.

The use of [C{\sc ii}] as an H$_2$ tracer is frequently adopted for high-$z$ galaxies \citep[e.g.][]{Zane18,Madd20,Vizg22}, thanks to ALMA's ability to observe this line in the very distant Universe. Using [C{\sc ii}] for such estimations is, however, not straight-forward since significant fraction of the total [C{\sc ii}] luminosity can originate from the non-molecular ISM, such as H{\sc ii} regions and the warm neutral medium \citep[e.g.][]{Accu17,Bisb22}. For the [C{\sc ii}] emission to correlate with $N$(H$_2$) for high columns of H$_2$, a heating mechanism is needed capable of exciting this line --such as cosmic-rays (see \citetalias{Bisb21})-- given its high excitation temperature of $h\nu/k\sim91.2\,{\rm K}$. Panels~(a)-(d) indicate a decreasing trend of $\cal X_{\rm [CII]}$ with increasing $\chi/\chi_0$ and $\zeta_{\rm CR}$. For metallicities of $\gtrsim0.5\,{\rm Z}_{\odot}$ it can vary up to two orders of magnitude depending on the combination of the ISM conditions. For high metallicities (e.g. $Z\sim2\,{\rm Z}_{\odot}$), [C{\sc ii}] shows a strong correlation for $\zeta_{\rm CR}\gtrsim10^{-15}\,{\rm s}^{-1}$. Such conditions are met in starburst galaxies in the high-$z$ Universe as well as the Galactic Center \citep{Give02,Clar13}.

The debate between [C{\sc i}](1-0) and CO(1-0) as `best of H$_2$ gas tracers' is of great interest (e.g. \citealt{Papa04,Bola13,Offn14} and \citetalias{Bisb21}). From the [C{\sc i}](1-0) results shown in panels~(e)-(h), it is found that its corresponding ratio with N(H$_2$) remains approximately constant under molecular conditions for $Z=0.1\,{\rm Z}_{\odot}$. For similar $\zeta_{\rm CR}$ and $\chi/\chi_0$ pairs (e.g. $\zeta_{\rm CR}\lesssim10^{-15}\,{\rm s}^{-1}$ and $\chi/\chi_0<5$) in all other metallicity cases, $\cal X_{\rm [CI](1-0)}$ also remains to within the same range, thus making it to weakly depend on $Z$. For $Z\gtrsim0.5\,{\rm Z}_{\odot}$ and as $\zeta_{\rm CR}$ increases, the latter ratio slightly decreases while for higher $\chi/\chi_0$, it slightly increases.

Panels~(i)-(l) illustrate the dependency of $\cal X_{\rm CO(1-0)}$ on the ISM parameters. There is a stronger correlation with metallicity in agreement with observations 
\citep[e.g.][]{Genz12,Shi15,Shi16,Amor16,Schr17}. In the $Z=0.1$ and $0.5\,{\rm Z}_{\odot}$ cases, $\cal X_{\rm CO(1-0)}$ has a minimum for $\zeta_{\rm CR}\sim10^{-15}\,{\rm s}^{-1}$ and low $\chi/\chi_0$. For higher metallicities (panels~k, l), it depends predominantly on the value of $\zeta_{\rm CR}$. In particular, $\cal X_{\rm CO(1-0)}$ can vary up to one order of magnitude as $\zeta_{\rm CR}$ increases. On the contrary, panels~(f)-(h) show that for high $\zeta_{\rm CR}$, $\cal X_{\rm [CI](1-0)}$ is approximately constant making it an excellent alternative tracer in such environments \citep{Bisb15,Bisb17b}. Overall, the pattern of the dominant carbon emitter illustrated in the bottom row of Fig.~\ref{fig:cf} is reflected in the $\cal X_{*}$ ratios of Fig.~\ref{fig:conversions}; the ratio of the dominant emitter minimizes when its antenna temperature peaks. 

\subsection{The HCO$^+$/CO abundance ratio as a cosmic-ray tracer}
\label{sssec:crs}

The abundance ratio of HCO$^+$/CO has been used in the past to infer the cosmic-ray ionization rate \citep{Woot79,Case98,Vaup14}. In particular, elevated HCO$^+$/CO abundance ratios have been connected with increased cosmic-ray ionization rates. \citet{Gach19} obtain high HCO$^+$/CO abundance ratios for models with high $\zeta_{\rm CR}$ values and without embedded sources at high column densities (their HNI model), which can change the local chemistry and thus affect that ratio.

\begin{figure}
    \centering
	\includegraphics[width=0.48\textwidth]{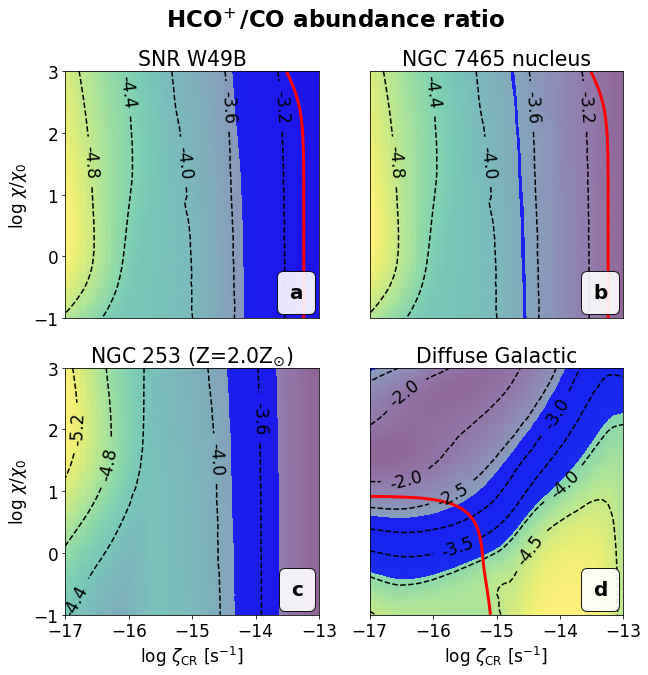}
    \caption{The HCO$^+$/CO abundance ratio. The blue-shadowed regions represent observations (see \S\ref{sssec:crs} for details). Panels~(a)-(c) show this ratio for the star-forming $A_{\rm V,obs}$-PDF. Panels~(a) and (b) are for $Z=1.0\,{\rm Z}_{\odot}$ while panel~(c) for $Z=2.0\,{\rm Z}_{\odot}$. In these cases, the HCO$^+$/CO abundance ratio depends almost entirely from the $\zeta_{\rm CR}$ value. Panel~(d) shows the ratio for the non-star-forming $A_{\rm V,obs}$-PDF for $Z=1.0\,{\rm Z}_{\odot}$. In this case and under molecular conditions, the abundance ratio does not depend strongly on $\zeta_{\rm CR}$ but rather on the $\chi/\chi_0$ value.}
    \label{fig:CRtracers}
\end{figure}

Figure~\ref{fig:CRtracers} shows HCO$^+$/CO abundance ratio maps for both $A_{\rm V,obs}$-PDF cases considered and how these compare with observations. Panels~(a)-(c) illustrate the HCO$^+$/CO ratio for the star-forming $A_{\rm V,obs}$-PDF for $Z=1.0\,{\rm Z}_{\odot}$ (panels~a, b) and for $Z=2.0\,{\rm Z}_{\odot}$ (panel~c). Panel~(d) shows the non-star-forming PDF for $Z=1.0\,{\rm Z}_{\odot}$.
In all cases, the observed data are marked in blue shadow. As can be seen, the star-forming PDFs show a strong dependence of HCO$^+$/CO ratio with $\zeta_{\rm CR}$. However, it is found that under molecular conditions and for the non-star-forming PDF, it depends more strongly on the value of $\chi/\chi_0$ rather the $\zeta_{\rm CR}$ one.

Panel~(a) highlights the observed HCO$^+$/CO abundance ratio in the supernova remnant W49B obtained by \citet{Zhou22} who reported a ratio in the range of $\log_{10}({\rm HCO^+/CO})=-3.52$ to $-2.70$. They attributed it to the presence of high $\zeta_{\rm CR}$ and suggested that it can well be more than two orders of magnitude higher than the average Galactic one. Indeed, this ratio fits for high $\zeta_{\rm CR}$ in our models indicating a $\zeta_{\rm CR}\gtrsim3\times10^{-15}\,{\rm s}^{-1}$. While \citet{Zhou22} do not report on the column density distribution, the selected $A_{\rm V,obs}$-PDF used here is to represent a dense clump.

Panel~(b) shows the observations of \citet{Youn21} for the nucleus of NGC~7465 which is a low-redshift early-type galaxy \citep{Capp11}. This panel has the same PDF and metallicity as panel~(a). The observed metallicity of this system is approximately solar. \citet{Youn21} calculated the $^{12}$CO and HCO$^+$ column densities for two representative temperatures (20 and 90~K) in the optically thin limit. These lead to a range of HCO$^+$/CO ratio of $\log_{10}{\rm HCO^+/CO}=-3.75$ to $-3.73$, for which {\sc PDFchem} finds a $\zeta_{\rm CR}\sim3-4\times10^{-15}\,{\rm s}^{-1}$. These elevated $\zeta_{\rm CR}$ values are reasonable to expect in an ISM region located in the center of galaxies such as NGC~7465.

Panel~(c) shows the observed ratio for the central part in the NGC~253 galaxy \citep{Krie20}. This central part contains several regions of proto-super star clusters and they are thus best represented with the star-forming $A_{\rm V,obs}$-PDF. NGC~253 is known to be a starburst galaxy of high star-forming activity \citep{Boll13b} and of super-solar metallicity \citep[e.g. $Z\sim2.19\,{\rm Z}_{\odot}$][]{Gall08}. The shadowed region in panel~(c) represents the mean and the $1\sigma$ standard deviation of the \citet{Krie20} reported values. These lead to a high $\zeta_{\rm CR}$ of approximately $10^{-14}\,{\rm s}^{-1}$. Interestignly, such high values of cosmic-ray ionization rate have been independently reported \citep[e.g.][]{Hold21}, confirming the {\sc PDFchem} findings. 

Finally, in panel~(d) the resultant calculations for a region representing the diffuse ISM of Milky Way are shown. Here, the non-star-forming $A_{\rm V,obs}$-PDF has been used for a better comparison. The observed values were taken by \citet{Goda14} and are shown in the aforementioned panel. Under molecular conditions, the reported ratio reveals an FUV intensity in the range of $\chi/\chi_0\sim1-5$, which matches with the average FUV intensity of Milky Way away from the Galactic Center. Under atomic conditions (which reveal a much higher pair of $\chi/\chi_0$ and $\zeta_{\rm CR}$), both lines are expected to be very weak and thus non-observable.

\subsection{Line ratios as FUV and $\zeta_{\rm CR}$ diagnostics}

Many lines in the carbon cycle ([C{\sc ii}], both [C{\sc i}] and high-$J$ CO transitions) have been observed at a redshift of $\sim2-3$ and beyond \citep[see relevant works in the reviews of][]{Hodg20,Wolf22}, their combinations of which can constrain the ISM environmental parameters using {\sc PDFchem}, thus offering a deeper understanding on the conditions leading to star-formation. While such observations have revealed the cosmic history of star-formation, they have yet to quantify the connection between the ISM environmental parameters and the SFR. The latter quantity is generally traced with the [C{\sc ii}] emission at high redshifts, since the [C{\sc ii}]-SFR relation has been extensively explored with observations \citep[e.g.][]{Stac91,Stac10,Bose02,DeLo11,Herr15} and simulations \citep[e.g.][]{Olse15,Lupi20,Bisb22}. It is reasonable to expect that high SFRs imply high $\zeta_{\rm CR}$ and possibly also high $\chi/\chi_0$ intensities, since both should be enhanced due to the expected higher rate of supernova explosions. In general, high SFRs are found in mergers, starbursts, ULIRGs and submillimeter galaxies \citep[e.g.][]{Smai97,Hugh98,Dadd10,Dadd15}.

The [C{\sc ii}]/[C{\sc i}](1-0) ratio explored in this work, shows that it may well be used to diagnose FUV intensities since its dependence on $\zeta_{\rm CR}$ is weak (for $\lesssim10^{-14}\,{\rm s}^{-1}$) regardless to the choice of $A_{\rm V}$-PDF \citep[see also][for an analysis with uniform density slabs]{Both17}. This remains valid even for different metallicities as can be seen in panel~(s) of Figs.~\ref{fig:lrt0p1}-\ref{fig:crt0p5}. Interestingly, for $Z=2\,{\rm Z}_{\odot}$ (panel~\ref{fig:crt2p0}s) the above ratio shows a stronger dependence on $\zeta_{\rm CR}$ for $\zeta_{\rm CR}\gtrsim10^{-15}\,{\rm s}^{-1}$. \citet{Both17} report a [C{\sc ii}]/[C{\sc i}](1-0) ratio in the (logarithmic) range of $\sim0.9-1.6$ for ten high-redshift ($z\sim4$) dusty star-forming galaxies and they further perform PDR modelling assuming an average $A_{\rm V,obs}\sim7\,{\rm mag}$. This matches with our star-forming PDF and from panel~(s) of Fig.~\ref{fig:crt} it is found that the above ratio predicts an FUV intensity of $\log_{10}\chi/\chi_0>2.5$. Interestingly, these high intensities calculated with {\sc PDFchem} are matching those estimated by \citet{Both17}.

\begin{figure}
    \centering
	\includegraphics[width=0.48\textwidth]{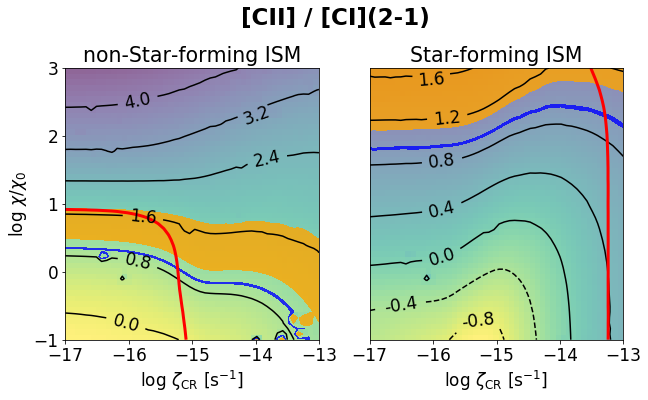}
    \caption{The average brightness temperature of [C{\sc ii}]/[C{\sc i}](2-1) for the non-star-forming ISM (left panel) and the star-forming ISM (right panel). The blue thick line and the orange shaded region correspond to the \citet{Vene17} and \citet{Pens21} observations, respectively.}
    \label{fig:ciic21}
\end{figure}

Due to the weak dependence on CRs, the [C{\sc ii}]/[C{\sc i}](2-1) line ratio has been considered as a tracer for the presence of X-rays in the ISM. This ratio is illustrated in Fig.~\ref{fig:ciic21}, which shows a strong dependence on the FUV intensity, like the [C{\sc ii}]/[C{\sc i}](1-0) explored above. Earlier models of \citet{Meij07} showed that this ratio depends strongly on the intensity of X-ray flux; it is increasing with increasing X-ray fluxes. This led \citet{Vene17} who studied the ISM of a very distant ($z\sim7.1$) quasar J1120+0641 to conclude that the aforementioned (lower limit) ratio of $\log_{10}(\rm [CII]/[CI](2-1))=0.94$
originates from existing PDRs that are a result of the FUV radiation from hot stars, rather than hard X-rays from the accreting central black hole. This observed ratio is marked with blue line in Fig.~\ref{fig:ciic21}. {\sc PDFchem} predicts an $\log\chi/\chi_0\sim0.5-2$ depending on the PDF distribution (for $Z=1.0\,{\rm Z}_{\odot}$. In addition, \citet{Pens21} used the [C{\sc ii}]/[C{\sc i}](2-1) ratios of $\log_{10}(\rm [CII]/[CI](2-1))\sim1.11-1.72$ to identify PDRs in the galaxies PJ231-20 ($z=6.23$) and PJ308-21 ($z=6.59$) which are two quasar host galaxies. The orange shadowed region of Fig.~\ref{fig:ciic21} marks these observations and as with the last finding, they may correspond to high FUV intensities for high-$A_{\rm V,obs}$ PDF distributions.

\citet{Boog20} studied the CO(2-1)/CO(1-0) and CO(3-2)/CO(1-0) ratios in a sample of 22 star-forming galaxies in $0.46<z<3.60$ as part of the ASPECS\footnote{ALMA Spectroscopic Survey in the Hubble
Ultra Deep Field} survey. They found a CO(2-1)/CO(1-0) ratio of $\sim0.75$ for $z<2$ and a CO(3-2)/CO(1-0) ratio of $\sim0.77$ for $z\geq2$, arguing that the higher excitation of the latter ratio is a result of the $\Sigma_{\rm SFR}$ increase with redshift. Interestingly, the CO(2-1)/CO(1-0) ratio fits our results for the star-forming $A_{\rm V,obs}-$PDF for a $\zeta_{\rm CR}$ of a few $\times10^{-15}\,{\rm s}^{-1}$ (panel~\ref{fig:crt}l), regardless to the FUV intensity, while for the CO(3-2)/CO(1-0) ratio we find a $\zeta_{\rm CR}$ of a few $\times10^{-14}\,{\rm s}^{-1}$ (panel~\ref{fig:crt}m). The higher $\zeta_{\rm CR}$ reflects the global picture in which we expect to find higher cosmic-ray energy densities in star-forming galaxies, particularly at $z\sim2-3$.

The pattern of CO(5-4)/CO(1-0) ratio illustrated in the panel~(\ref{fig:crt}o), shows that high-$J$ CO lines depend very weakly on the FUV intensity and thus can be used to constrain heating processes that operate at high column densities, other than stellar photons. In this regard, a combination of CO SLEDs with dust continuum spectral energy distributions \citep[e.g.][]{Harr21} can reveal these heating processes as to whether or not they are photon-dominated, by estimating the so-called $Y$-factor ($Y=\Gamma_{\rm PE}/\Lambda_{\rm tot}$, the ratio of photoelectric heating to the total cooling, see \citealt{Papa14}). The advent of ALMA has made possible to observe high-$J$ CO lines in the distant Universe which points to the existence of warm and dense medium. For instance, the $J=10-9$ transition requires a gas temperature of $T_{\rm gas}\gtrsim300\,{\rm K}$ to be excited and has a critical density of $n_{\rm crit}\sim10^6\,{\rm cm}^{-3}$. In addition, the NOEMA detection of very high-$J$ CO lines ($J=17-16$) in very distant ($z>6$) galaxies \citep{Gall14} makes it possible to investigate the ISM and constrain its parameters even in the Early Universe. In these galaxies, it is not only cosmic-rays that may cause an elevated SLED. X-rays \citep[e.g.][]{Vall18,Vall19,Pens21}, shock heating \citep[e.g.][]{Li20,Riec21} and/or turbulence \citep{Harr21} can also play a significant role. Exploring high-$J$ transitions with the presented version of {\sc PDFchem} can help the investigation of these heating processes, primarily by disentangling the contribution due to FUV photons. Future versions of {\sc PDFchem} will be able to provide a further constrain on the heating processes. 

In general, for applications of {\sc PDFchem} in high-redshift galaxies in which the $A_{\rm V,obs}$-PDFs remain largely unknown, we recommend the use of distributions corresponding to a diffuse or a dense medium according to local observations \citep[e.g.][]{Spil21,Ma22} or inspired from simulations \citep[e.g. see recent works of][]{Velt19,Seif20b,Olse21,Appe22}. These can make possible to obtain an insight of the PDR trends and to provide a probable range of environmental parameters which, using additional forward modelling, may identify the ISM conditions. 

\section{Conclusions}
\label{sec:conclusions}

In this paper we present the {\sc PDFchem} algorithm, which continues the approach of \citetalias{Bisb19} in determining the atomic and molecular mass content of the ISM using probability distributions of physical parameters. For the purposes of PDR simulations, we consider a variable density distribution resulting from the $A_{\rm V,eff}-n_{\rm H}$ relationship which replaces the grid of uniform density slabs used in \citetalias{Bisb19}. We have performed radiative transfer calculations to determine ratios between emissions of the most commonly used lines of the carbon cycle. The grid of PDR simulations consists of 6,400 one-dimensional runs covering a wide range of cosmic-ray ionization rates, FUV intensities and metallicities. We applied {\sc PDFchem} to two hypothetical $A_{\rm V,obs}$-PDFs representing a non-star-forming and a star-forming ISM distributions. Our results can be summarized as follows:

\begin{enumerate}
    \item The demanding inhomogeneous PDR calculations in advanced three-dimensional hydrodynamical simulations can be reproduced by a simple one-dimensional density distribution (the `variable density slab') and which results from the empirical $A_{\rm V,eff}-n_{\rm H}$ relation of Eqn.~\ref{eq:aveff_nh}. We distinguish between the effective (local) and the observed visual extinctions, $A_{\rm V,eff}$ and $A_{\rm V,obs}$ respectively, and we show that for a better interpretation of the observed PDR data, a conversion relation between these two quantities must be considered (Eqn.~\ref{eq:aveff_avobs}). 
    \item The non-star-forming distribution remains molecular for a combination of low FUV intensities and low $\zeta_{\rm CR}$. In the extreme case of $Z=0.1\,{\rm Z}_{\odot}$ it is always atomic, whereas in the case of $Z=2\,{\rm Z}_{\odot}$ is remains molecular for higher combinations of $\chi/\chi_0$ and $\zeta_{\rm CR}$. Its molecular phase is almost always C{\sc ii}-dominated in terms of the abundance, and it is [C{\sc ii}]-bright for $\chi/\chi_0>1$. For lower FUV intensities it is either CO(1-0) or [C{\sc i}](1-0) bright depending on $\zeta_{\rm CR}$. This distribution remains in general optically thin in [C{\sc ii}] and [C{\sc i}](1-0), but can become quickly CO(1-0) optically thick.
    \item The star-forming distribution remains molecular for almost any combination of $\chi/\chi_0$--$\zeta_{\rm CR}$ explored, except when $Z=0.1\,{\rm Z}_{\odot}$ for which it is atomic for $\chi/\chi_0>5$ and $\zeta_{\rm CR}>10^{-15}\,{\rm s}^{-1}$. In terms of the abundance, the molecular phase is dominated either by C{\sc ii}, C{\sc i} or CO depending on the ISM conditions. Similarly, it can be either [C{\sc ii}], [C{\sc i}](1-0) or CO(1-0) bright when it comes to line emission. In general, the power-law tail does not significantly affect the results of the log-normal component, but we do observe a change in the conditions leading to the H{\sc i}-to-H$_2$ transition. This distribution remains in principle optically thin in [C{\sc ii}]. It is also optically thin in [C{\sc i}](1-0) for sub-solar metallicities only. As with the non-star-forming distribution, it is almost always optically thick in CO(1-0).
    \item We find that the CO-to-H$_2$ conversion factor depends more strongly on the ISM environmental parameters than the CI-to-H$_2$ does. The latter is found to depend weakly on metallicity and remains approximately constant for $\zeta_{\rm CR}\gtrsim10^{-15}\,{\rm s}^{-1}$, making it a powerful alternative tracer to CO in such environments. As an H$_2$ tracer, [C{\sc ii}] is found to remain remarkably independent of metallicity. It decreases when both the FUV intensity and metallicity increase and for high cosmic-ray energy densities, it depends almost entirely on $\zeta_{\rm CR}$.
    \item We identify the HCO$^+$/CO abundance ratio as a tool to constrain the $\zeta_{\rm CR}$ value. Contrary to the non-star-forming distribution, we find that for the star-forming $A_{\rm V,obs}$-PDF, it remains largely independent of the value of the FUV intensity. We also identify the [C{\sc ii}]/[C{\sc i}](1-0) and [C{\sc ii}]/[C{\sc i}](2-1) line ratios as useful tools to constrain the FUV intensity, as they both depend weakly on the $\zeta_{\rm CR}$ value.
\end{enumerate}

The significant advantage of {\sc PDFchem} over 3D-hydro+radiative transfer codes to solve for the PDR properties in a negligible computational time, allows the exploration of various combinations of PDFs and to understand the ISM properties of large-scale systems including local clouds and high-redshift galaxies.

\section*{Data availability}

The data underlying this article will be shared on reasonable request to the corresponding author.

\section*{Acknowledgements}

The authors thank the anonymous referee for useful comments which improved the clarity of this work. The authors thank Pierre N\"urnbenger for improving the quality of Figures~\ref{fig:lf} and \ref{fig:cf} and Kei Tanaka for the discussions. TGB acknowledges support from Deutsche Forschungsgemeinschaft (DFG) grant No. 424563772. EvD is supported by the European Research Council (ERC) under the European Union's Horizon 2020 research and innovation program (grant agreement No. 101019751 MOLDISK). CYH acknowledges support from the DFG via German-Israel Project Cooperation grant STE1869/2-1 GE625/17-1. 
%%%%%%%%%%%%%%%%%%%%%%%%%%%%%%%%%%%%%%%%%%%%%%%%%%



%\bibliographystyle{mnras}


%%%%%%%%%%%%%%%%%%%%%%%%%%%%%%%%%%%%%%%%%%%%%%%%%%

%%%%%%%%%%%%%%%%% APPENDICES %%%%%%%%%%%%%%%%%%%%%

\appendix

\section{Construction of the variable density slab}
\label{app:slab}

\begin{figure}
    \centering
    \includegraphics[width=0.48\textwidth]{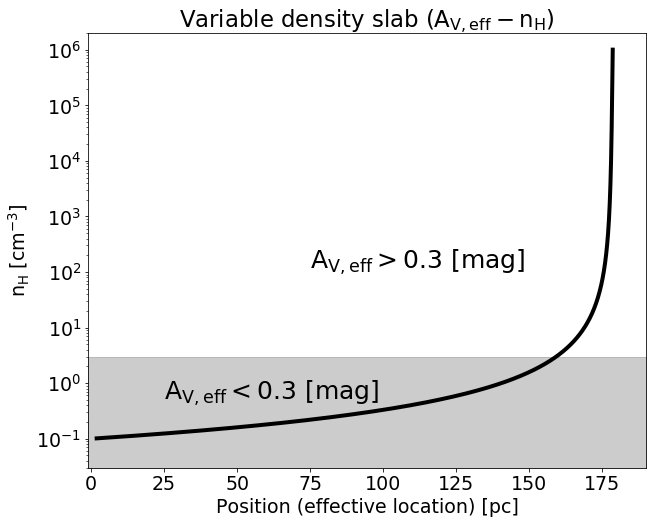}
    \caption{Density profile of the variable density slab resulting from the $A_{\rm V,eff}-n_{\rm H}$ relation. The position ($x$-axis) describes the effective location of each $n_{\rm H}$ in the three-dimensional space and does not represent the size of the cloud. The shadowed region has $A_{\rm V,eff}<0.3\,{\rm mag}$ and it is atomic for average (Milky-Way) ISM environmental conditions.}
    \label{fig:varslab}
\end{figure}

To construct the variable density distribution described in \S\ref{ssec:avenh}, we consider a minimum density of $n_\mathrm{min}=0.1\,\mathrm{cm}^{-3}$ and a maximum density of $n_\mathrm{max}=10^6\,\mathrm{cm}^{-3}$. The resolution is taken to be $30$ points logarithmic spaced per $n$~dex, thus a total of $210$ points. The difference between two consecutive densities corresponds to a $\Delta x$ displacement satisfying the relation
\begin{eqnarray}
\label{eqn:deltax}
\Delta x(n_i) = \frac{0.05\left(e^{1.6n_i^{0.12}}-e^{1.6n_{i-1}^{0.12}}\right)}{A_\mathrm{V,0}n_i},&i\ge1.
\end{eqnarray}
In the above, $A_{\rm V,0}=6.3\times10^{-22}\,{\rm mag}\,{\rm cm}^{2}$ and $n_i$ is in units of $\rm cm^{-3}$, thus $\Delta x$ is in units of $\rm cm$. The special case of $i=0$ defines the edge of the cloud in which $\Delta x(n_0)=0$ for $n_0 \equiv n_\mathrm{min}$. Thus, the effective length of the cloud is $L=\sum_i\Delta x(n_i)\simeq178.56\,{\rm pc}$. This length represents the effective location of each $n_{\rm H}$ in the three-dimensional space and shall not be considered as the size of the cloud. Figure~\ref{fig:varslab} shows the resultant density profile in black solid line. Note that the slope of the function becomes very steep for $L\gtrsim175\,{\rm pc}$ corresponding to $n_{\rm H}\gtrsim10^2\,{\rm cm}^{-3}$.

\section{Comparison with radex}
\label{app:radex}

We perform a benchmarking test to compare our radiative transfer algorithm presented in \S\ref{ssec:rt} and used in {\sc PDFchem}, with the widely used {\sc radex}  algorithm\footnote{https://home.strw.leidenuniv.nl/$\sim$moldata/radex.html} \citep{vdTak07}. For this test we perform a {\sc 3d-pdr} calculation of an one-dimensional uniform density cloud with $n_{\rm H}=10^3\,{\rm cm}^{-3}$ H-nucleus number density and a constant temperature of $T_{\rm gas}=40\,{\rm K}$ everywhere in the cloud (isothermal). The cloud has an $A_\mathrm{V}=10^2\,\mathrm{mag}$ visual extinction. In the PDR calculations, we neglect the FUV radiation field by setting $\chi/\chi_0=0$ and we adopt a microturbulent velocity of $v_{\rm turb}=1\,{\rm km}\,{\rm s}^{-1}$. The latter corresponds to a {\sc radex} linewidth equal to
\begin{eqnarray}
\mathrm{Linewidth} = 2\sqrt{2\ln(2)}\sqrt{\frac{k_\mathrm{B}T_\mathrm{gas}}{m_\mathrm{mol}}+\frac{v_\mathrm{turb}^2}{2}}\simeq1.7\,\mathrm{km}\,\mathrm{s}^{-1},
\end{eqnarray}
where $k_\mathrm{B}$ is Boltzmann's constant and $m_\mathrm{mol}$ is the molecular mass of each coolant. 

\begin{figure}
    \centering
	\includegraphics[width=0.4\textwidth]{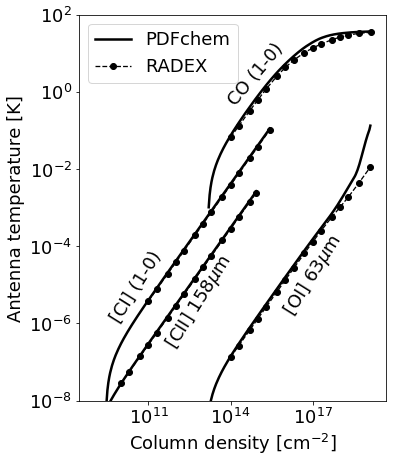}
    \caption{Antenna temperatures versus column density for CO~(1-0), [C{\sc i}]~(1-0), [C{\sc ii}]~158$\mu$m, [O{\sc i}]~$63\mu$m. Solid lines are the results of the \S\ref{ssec:rt} radiative-transfer scheme used in {\sc PDFchem}, and dashed-thick dotted line are {\sc radex} results. There is an excellent agreement between the two algorithms.}
    \label{fig:radex}
\end{figure}

We solve the radiative transfer equation and obtain the radiation temperatures, $T_\mathrm{r}$, of [C{\sc ii}]~$158\mu$m, [C{\sc i}]~(1-0), CO~(1-0) and [O{\sc i}]~$63\mu$m. Figure~\ref{fig:radex} shows the comparison between the radiative transfer scheme (``{\sc PDFchem}") and {\sc radex}. As can be seen the results obtained by both codes are in broad agreement. Similar behaviour is observed also with the rest of cooling lines we consider. 

\section{Effect of the suprathermal switch}
\label{sssec:supra}

As described in \S\ref{ssec:pdr}, the PDR calculations used for the above results include the switch of CO formation path via CH$^+$. It is interesting to explore the effect it has compared to the classical PDR approach \citep[e.g.][]{Roll07} which excludes it. For this, we have re-run the grid of models for $Z=1\,{\rm Z}_{\odot}$ and have applied our method to the non star-forming $A_{\rm V,obs}-$PDF case. We compare the H{\sc i}-to-H$_2$ transition and explore the pairs of $(\zeta_{\rm CR},\chi/\chi_0)$ that make the antenna temperature of CO $J=1-0$ the dominant one amongst all carbon phases.

\begin{figure}
    \centering
	\includegraphics[width=0.48\textwidth]{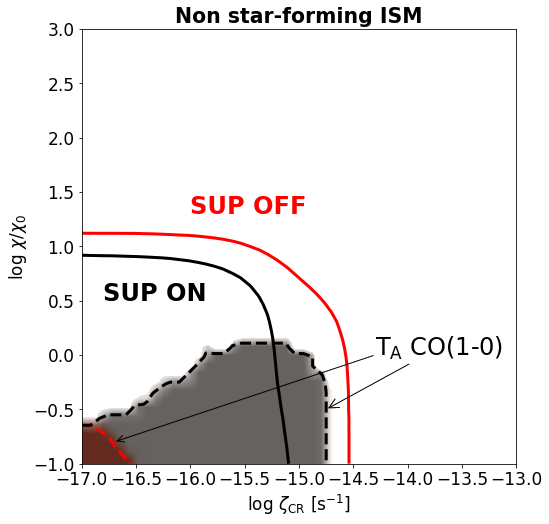}
    \caption{The H{\sc i}-to-H$_2$ transition and the antenna temperature of CO $J=1-0$ at solar metallicity for the non star-forming ISM distribution, when the suprathermal switch is enabled (`SUP ON', black lines) and disabled (`SUP OFF', red lines). The H{\sc i}-to-H$_2$ transition (solid lines) occurs at higher visual extinctions due to the destruction of H$_2$ by C$^+$. The shaded regions indicate when the antenna temperature of CO $J=1-0$ dominates over that of other species in the carbon cycle (gray coloured for `SUP ON', red coloured for `SUP OFF').}
    \label{fig:tr_sup}
\end{figure}

Figure~\ref{fig:tr_sup} shows this comparison when the suprathermal switch is enabled (`SUP ON'; black lines and gray region) and disabled (`SUP OFF'; red lines and region). Once the suprathermal switch is disabled, the H{\sc i}-to-H$_2$ is found to occur for higher pairs of $\chi/\chi_0$ and $\zeta_{\rm CR}$, which means that the transition occurs at higher depths into the cloud. This is due to the destruction of H$_2$ by C$^+$ in addition to photodissociation \citepalias[see also Appendix~A of][]{Bisb19}. The gray shadowed region, showing the range in which $T_{\rm A}$(CO~1-0) dominates, is much more extended in the `ON' case than in the `OFF' case, the latter marked in red shadow. In particular it is found that $T_{\rm A}$CO(1-0) dominates in the `OFF' case only for a very small range of ISM parameters with $\zeta_{\rm CR}\lesssim2\times10^{-16}\,{\rm s}^{-1}$ and $\chi/\chi_0\lesssim0.3$ as opposed in the `ON' case where $\zeta_{\rm CR}\lesssim2\times10^{-15}\,{\rm s}^{-1}$ and, in general, $\chi/\chi_0\lesssim1$. 

As a result, the suprathermal formation of CO via CH$^+$ can affect both the H{\sc i}-to-H$_2$ transition but most importantly it can make an ISM distribution to be bright in CO(1-0) for a much wider range of ISM environmental parameters.

\section{Optical depths for the non-star forming distribution}
\label{sssec:tau_nSF}

\begin{figure}
    \centering
    \includegraphics[width=0.48\textwidth]{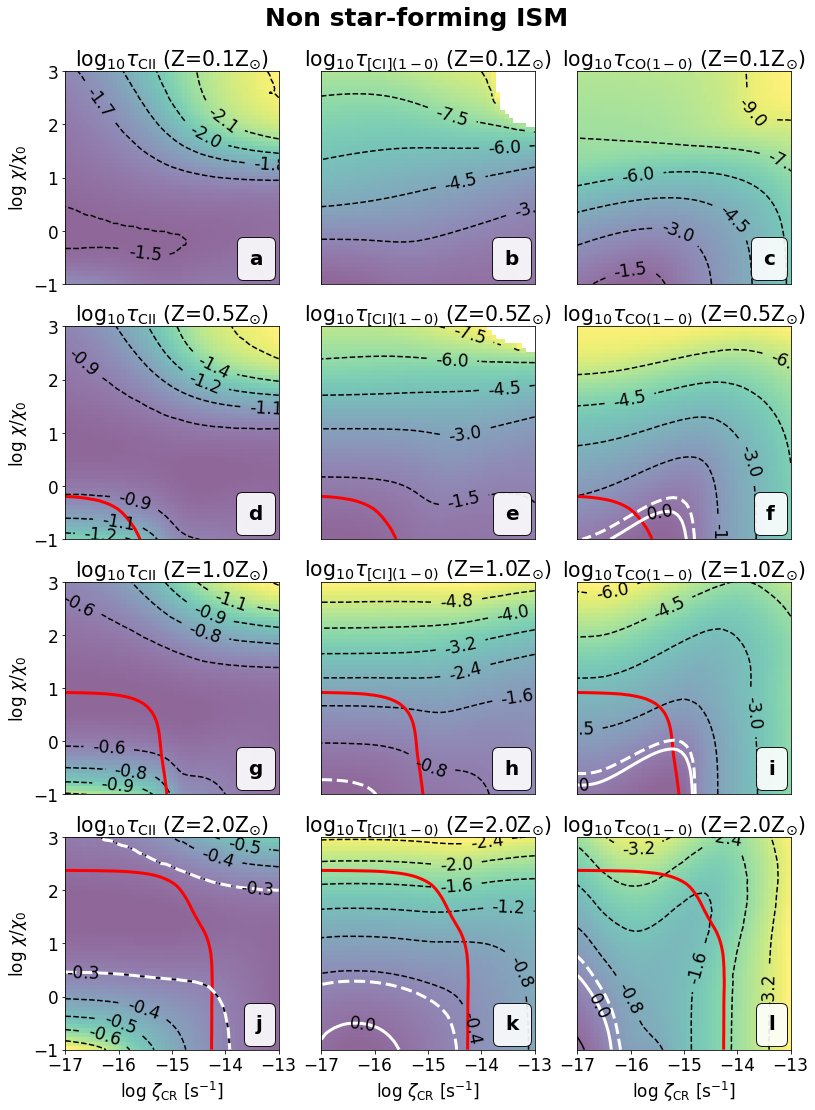}
    \caption{Optical depths of [C{\sc ii}]~158$\mu$m (left column), [C{\sc i}](1-0) (middle column) and CO(1-0) (right column) for $Z=0.1$, 0.5, 1.0 and $2.0\,{\rm Z}_{\odot}$ (top-to-bottom), for the non-star-forming distribution. The red solid line marks the H{\sc i}-to-H$_2$ transition. The white dashed and the white solid lines correspond to $\tau=0.5$, i.e. when optical depth effects start to become important, and $\tau=1$, i.e. when the line has become optically thick, respectively.}
    \label{fig:lt}
\end{figure}

Figure~\ref{fig:lt} shows optical depth ($\tau$) maps of the [C{\sc ii}]~$158\mu$m, [C{\sc i}](1-0) and CO(1-0) for all ISM environmental parameters. As with Fig.~\ref{fig:lrt}, the red solid line represent the H{\sc i}-to-H$_2$ transition. We also mark with white dotted lines the condition at which $\tau=0.5$ and with white solid lines the $\tau=1$ condition. These conditions show when optical depth effects start to be important and when the line has become optically thick, respectively. 

As discussed in \S\ref{sssec:cphase1}, for $Z=0.1\,{\rm Z}_{\odot}$ the distribution is largely atomic and [C{\sc ii}]~$158\mu$m bright, thus [C{\sc i}](1-0) and CO(1-0) are expected to be quite faint. Nevertheless, it is found that all aforementioned lines are very optically thin under all combinations of $\zeta_{\rm CR}$ and $\chi/\chi_0$ (panels~a-c). For $Z=0.5\,{\rm Z}_{\odot}$, the molecular phase of the distribution is also optically thin in [C{\sc ii}] and [C{\sc i}](1-0) (panels~d-e), but becomes optically thick in the CO(1-0) line as $\zeta_{\rm CR}$ approaches ${\sim}10^{-16}\,{\rm s}^{-1}$ (panel~f). 

For $Z=1\,{\rm Z}_{\odot}$, it is found that both [C{\sc ii}] and [C{\sc i}](1-0) remain optically thin (panels~g-h) with the latter to have $\tau{\simeq}0.5$ for only a combination of very low $\zeta_{\rm CR}$ and $\chi/\chi_0$. On the other hand, CO(1-0) (panel~i) becomes optically thick for $\chi/\chi_0\lesssim0.5$ when $\zeta_{\rm CR}\lesssim10^{-16}\,{\rm s}^{-1}$. Notably, for moderate cosmic-rays ($\zeta_{\rm CR}\sim7\times10^{-16}\,{\rm s}^{-1}$) it remains optically thick for FUV intensities up to $\chi/\chi_0\simeq1$. 

For $Z=2\,{\rm Z}_{\odot}$, $\tau_{\rm CII}\gtrsim0.5$ for $\chi/\chi_0\gtrsim5$ regardless of the $\zeta_{\rm CR}$ parameter for molecular conditions (panel~j). We, therefore, find that this line may experience optical depth effects especially at higher FUV intensities. The [C{\sc i}](1-0) line becomes optically thick, with $\tau_{\rm CI}{\sim}1$ for $\chi/\chi_0<0.5$ and for $\zeta_{\rm CR}\lesssim10^{-16}\,{\rm s}^{-1}$ (panel~k). Interestingly, CO(1-0) is optically thick for conditions similar to those for the [C{\sc i}](1-0) (panel~l). Looking at the response of $\tau_{\rm CO}$ for $Z=0.5$ and $1.0\,{\rm Z}_{\odot}$ (panels~f \& i), one would expect to find a distribution that would be much more optically thick in CO(1-0) for $Z=2.0\,{\rm Z}_{\odot}$. However, because higher metallicity implies higher visual extinction, the suprathermal switch plays a minor role here and ceases to be important at lower $A_{\rm V}$ than in the above cases. Consequently, CO does not form as quickly at low $A_{\rm V}$ as for $Z\lesssim1\,{\rm Z}_{\odot}$ and given that the non-star-forming ISM is a low column density distribution, reduces the optical depth of CO(1-0). It is further noted that should the $Z=0.1-1\,{\rm Z}_{\odot}$ models not include the suprathermal CO formation path, it would result in optically thin CO(1-0).

\section{Optical depths for the star-forming distribution}
\label{sssec:tau_SF}

\begin{figure}
    \centering
	\includegraphics[width=0.48\textwidth]{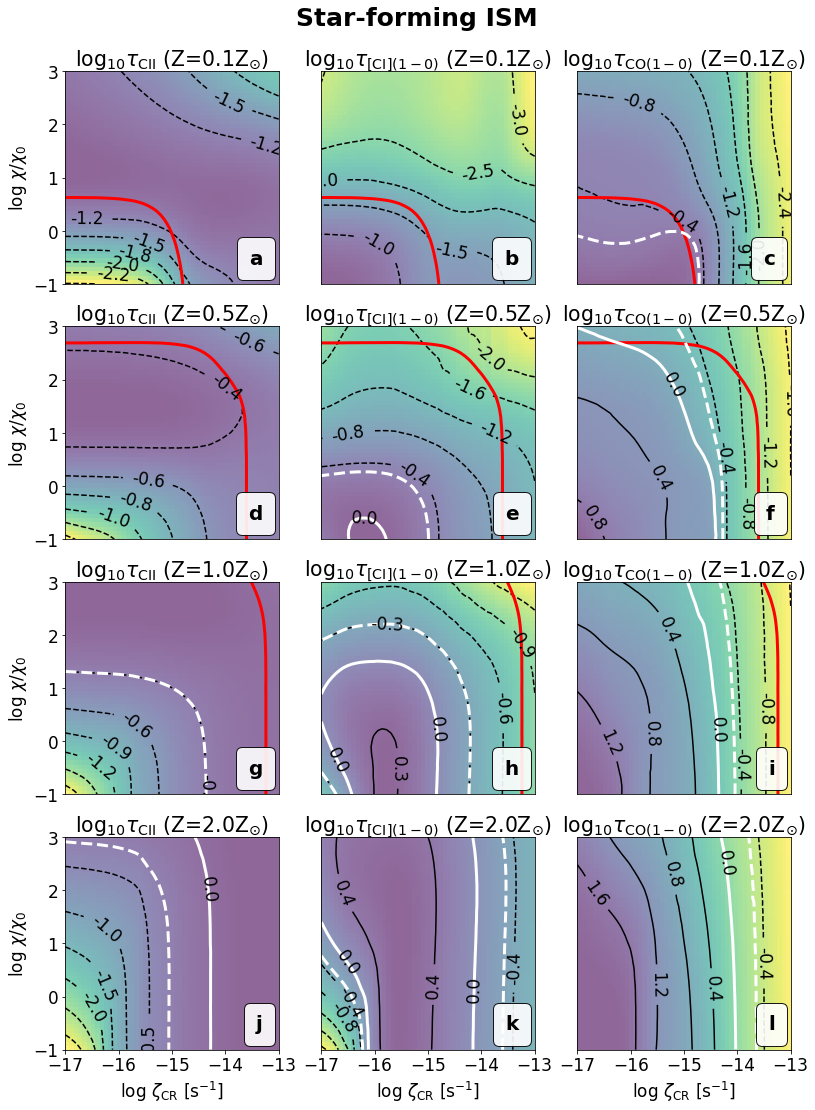}
    \caption{As in Fig.~\ref{fig:lt} for the star-forming ISM.}
    \label{fig:ct}
\end{figure}

Figure~\ref{fig:ct} shows the optical depths of the basic carbon cycle emission lines. For $Z=0.1\,{\rm Z}_{\odot}$, both [C{\sc ii}] and [C{\sc i}](1-0) remain optically thin (panels~a,b), while CO(1-0) has $\tau_{\rm CO}{\sim}0.5$ for molecular conditions and for $\chi/\chi_0\lesssim1$ (panel~c). For $Z=0.5\,{\rm Z}_{\odot}$, [C{\sc ii}] remains optically thin (panel~d) but [C{\sc i}](1-0) experiences optical depth effects for low FUV intensities and low $\zeta_{\rm CR}$ values (panel~e). On the other hand, CO(1-0) (panel~f) becomes quickly optically thick for $\zeta_{\rm CR}\lesssim10^{-15}\,{\rm s}^{-1}$ and, interestingly, $\tau_{\rm CO}$ depends primarily on the value of the cosmic-ray ionization rate. 

For $Z=1\,{\rm Z}_{\odot}$, [C{\sc ii}] experiences optical depth effects for $\zeta_{\rm CR}\lesssim8\times10^{-15}\,{\rm s}^{-1}$ and $\chi/\chi_0\gtrsim20$, as well as for $\zeta_{\rm CR}\gtrsim8\times10^{-15}\,{\rm s}^{-1}$ regardless to the FUV intensity (panel~g). For moderate $\zeta_{\rm CR}$ and for $\chi/\chi_0\lesssim10$, [C{\sc i}](1-0) becomes optically thick (panel~h). As in the $Z=0.5\,{\rm Z}_{\odot}$ case, CO(1-0) becomes optically thick for $\zeta_{\rm CR}{\lesssim}8\times10^{-15}\,{\rm s}^{-1}$ (panel~i). Finally, for $Z=2\,{\rm Z}_{\odot}$, [C{\sc ii}] is optically thick for $\zeta_{\rm CR}\gtrsim10^{-14}\,{\rm s}^{-1}$, and both [C{\sc i}](1-0) and CO(1-0) (panels~k,l) are also very optically thick for almost all combinations of $\zeta_{\rm CR}$--FUV intensities. 

As a result, for such an $A_{\rm V,obs}$-PDF, CO(1-0) is optically thick for almost all conditions explored while [C{\sc i}](1-0) and [C{\sc ii}] become optically thick for metallicities of $Z\gtrsim1.0\,{\rm Z}_{\odot}$.

\section{Abundances and line ratios for different metallicities}
\label{app:abnd}

\begin{figure*}
    \centering
	\includegraphics[width=0.98\textwidth]{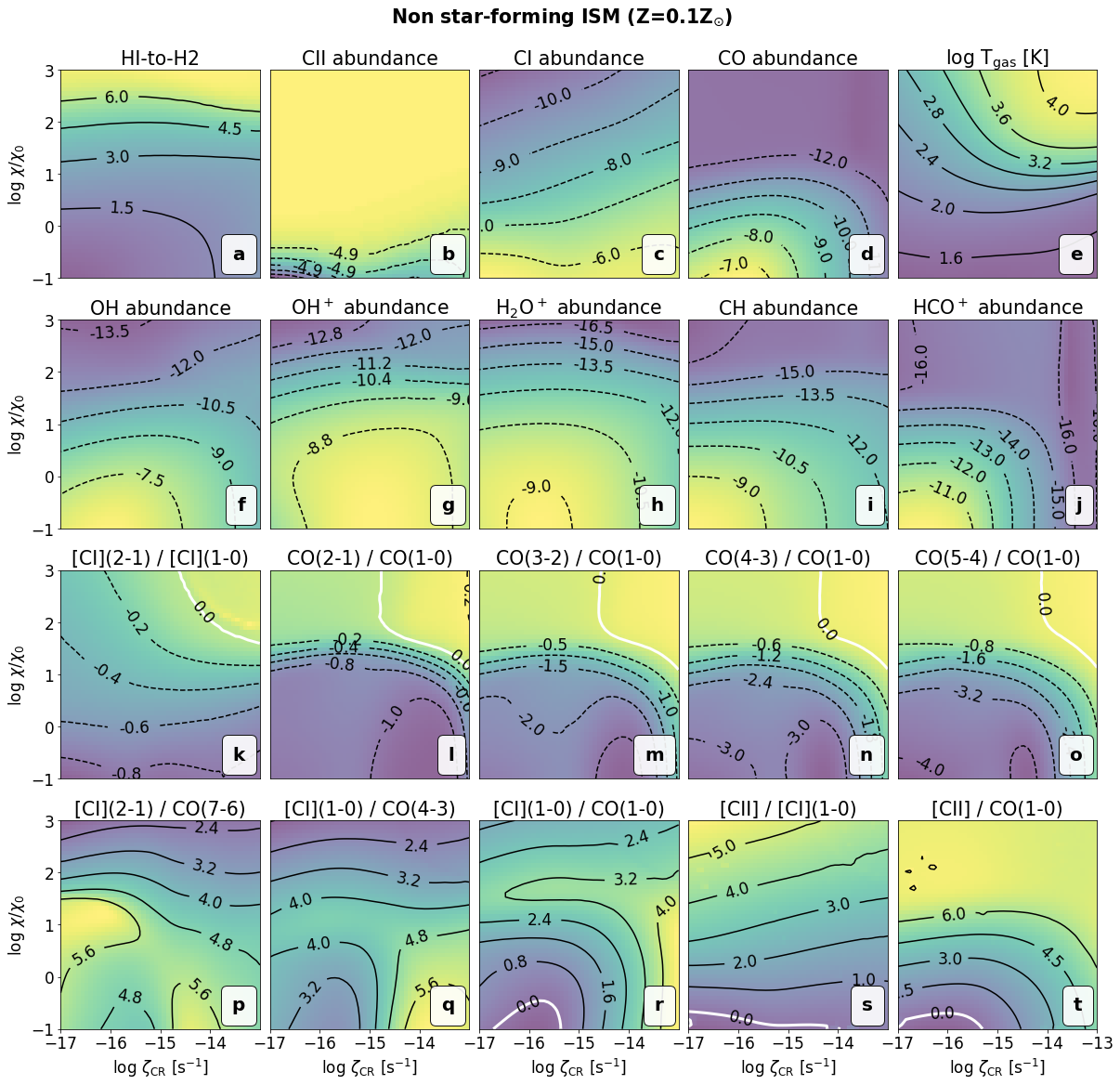}
    \caption{As Fig.\ref{fig:lrt} for $Z=0.1\,{\rm Z}_{\odot}$. Note that the medium is atomic in all cases.}
    \label{fig:lrt0p1}
\end{figure*}

\begin{figure*}
    \centering
	\includegraphics[width=0.98\textwidth]{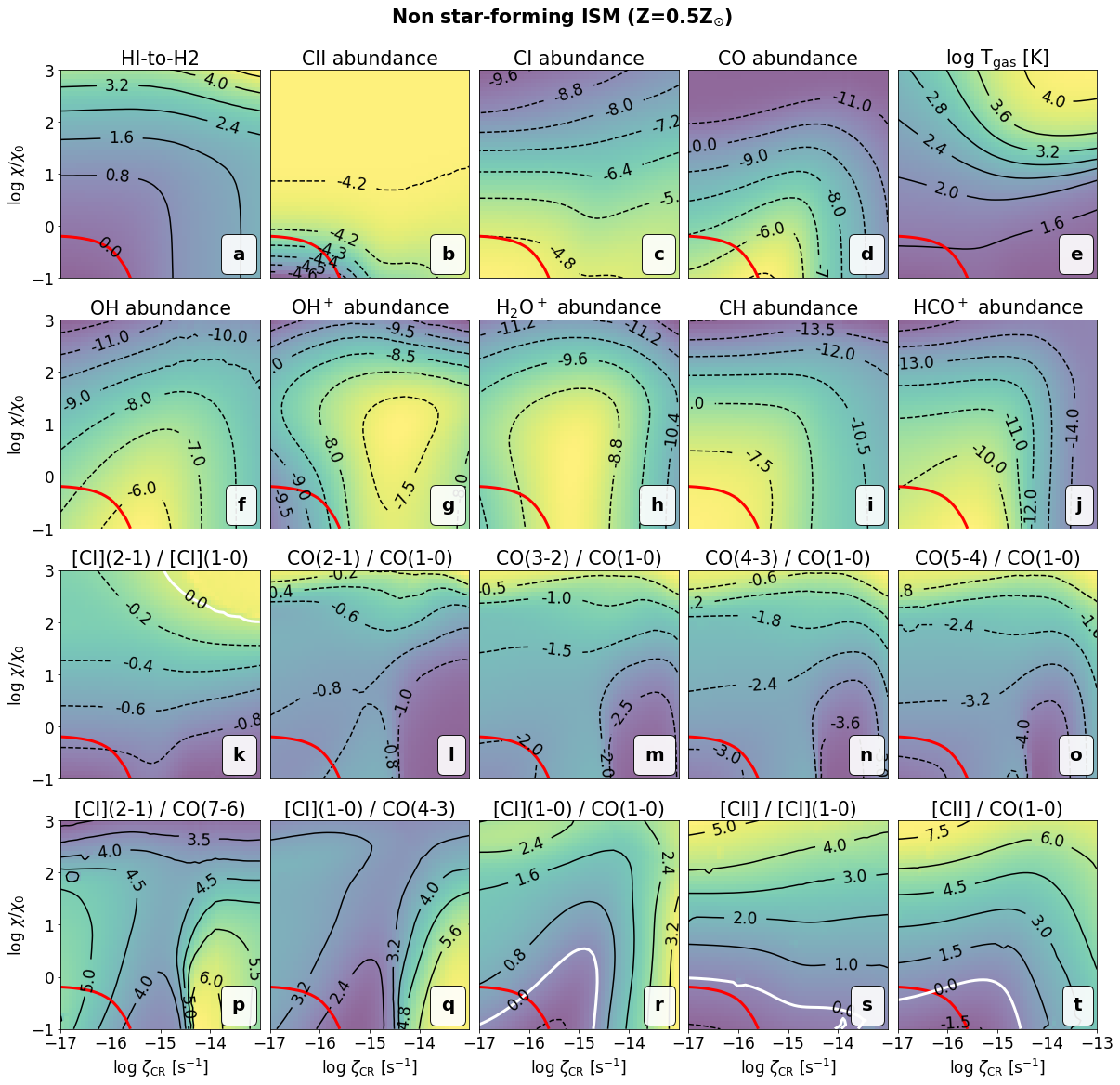}
    \caption{As Fig.\ref{fig:lrt} for $Z=0.5\,{\rm Z}_{\odot}$.}
    \label{fig:lrt0p5}
\end{figure*}

\begin{figure*}
    \centering
	\includegraphics[width=0.98\textwidth]{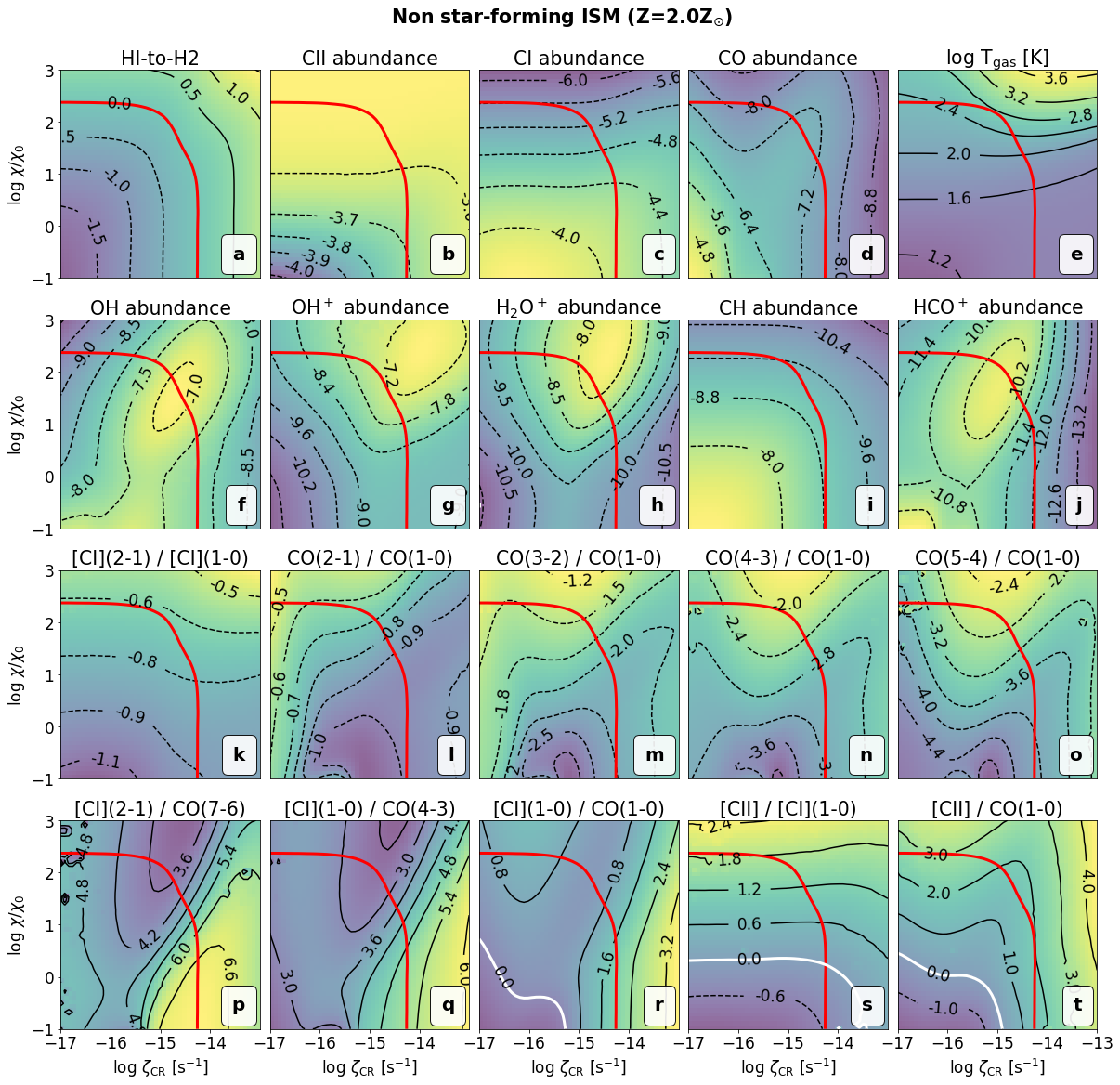}
    \caption{As Fig.\ref{fig:lrt} for $Z=2.0\,{\rm Z}_{\odot}$.}
    \label{fig:lrt2p0}
\end{figure*}

\begin{figure*}
    \centering
	\includegraphics[width=0.98\textwidth]{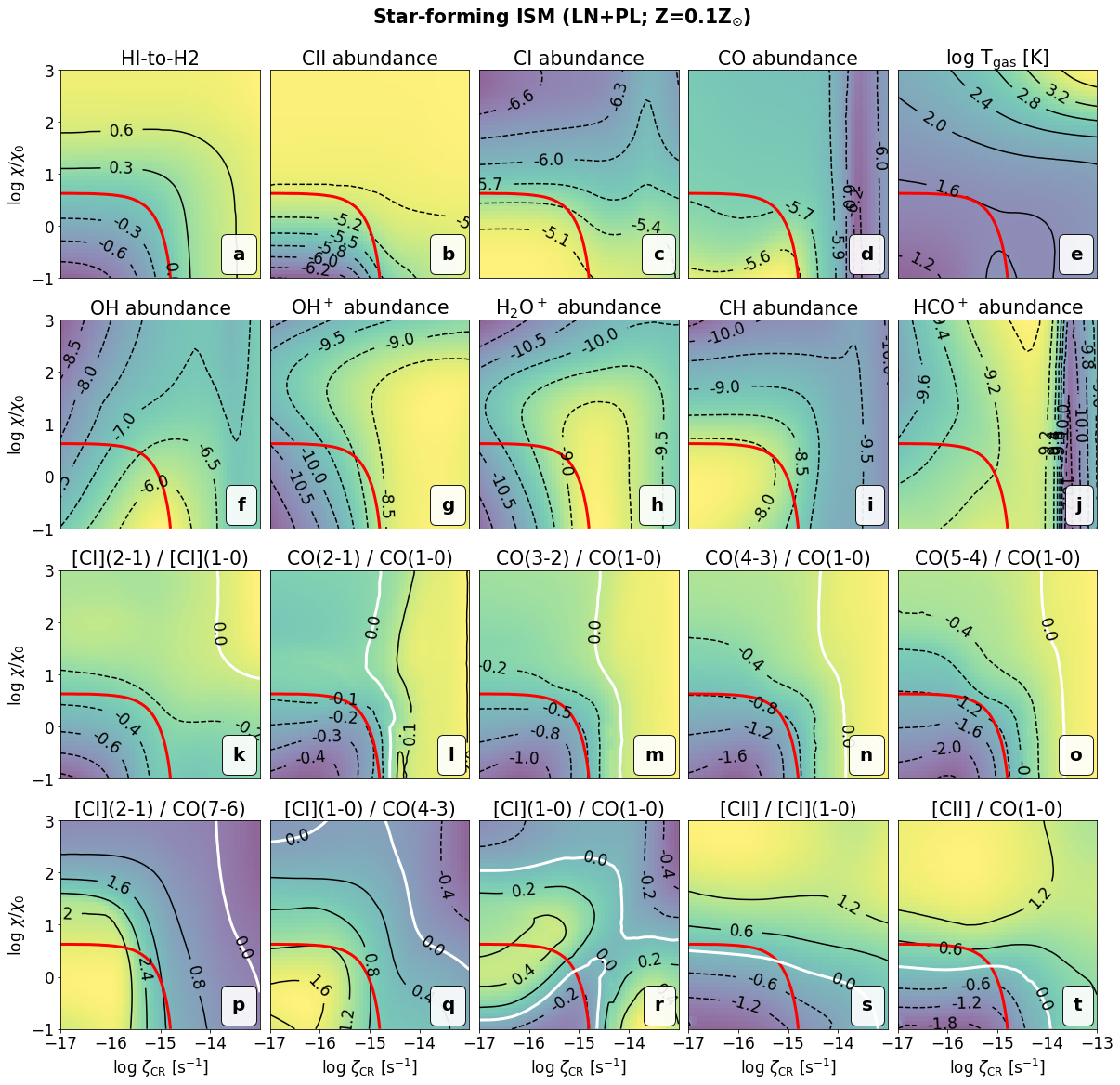}
    \caption{As Fig.\ref{fig:crt} for $Z=0.1\,{\rm Z}_{\odot}$.}
    \label{fig:crt0p1}
\end{figure*}

\begin{figure*}
    \centering
	\includegraphics[width=0.98\textwidth]{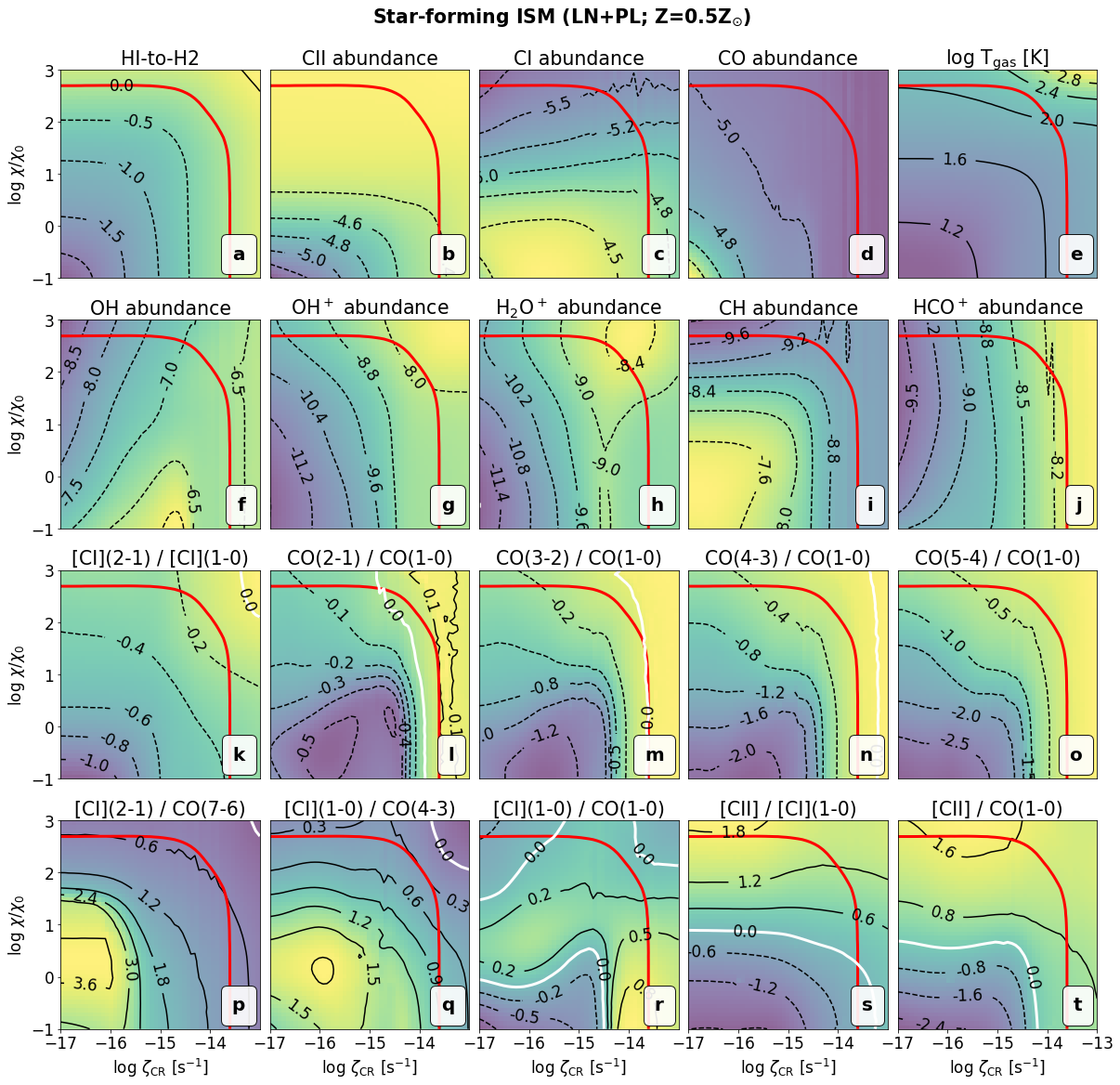}
    \caption{As Fig.\ref{fig:crt} for $Z=0.5\,{\rm Z}_{\odot}$.}
    \label{fig:crt0p5}
\end{figure*}

\begin{figure*}
    \centering
	\includegraphics[width=0.98\textwidth]{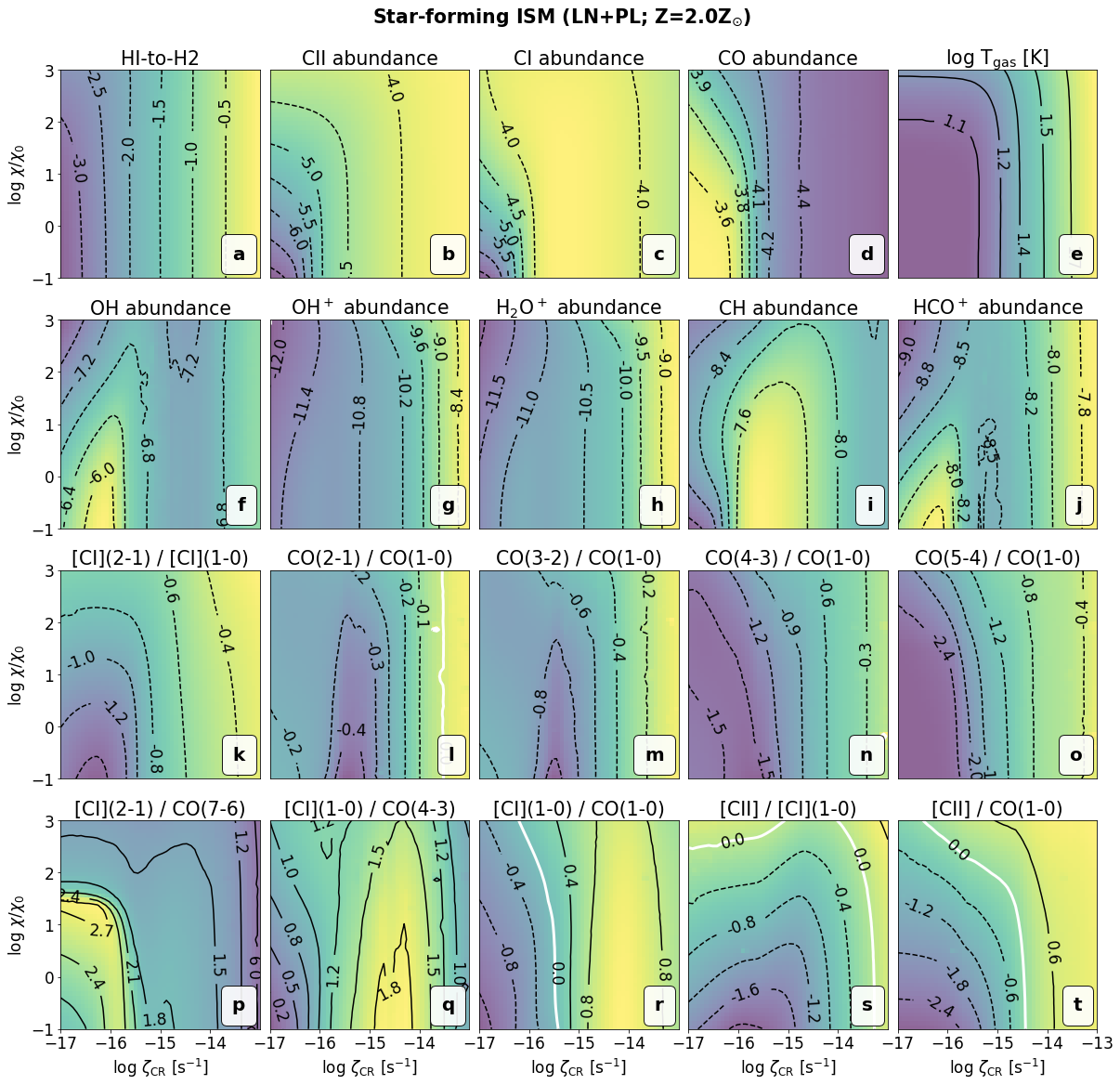}
    \caption{As Fig.\ref{fig:crt} for $Z=2.0\,{\rm Z}_{\odot}$. Note that the medium is molecular in all cases.}
    \label{fig:crt2p0}
\end{figure*}

Figures~\ref{fig:lrt0p1}-\ref{fig:lrt2p0} show the results for the non-star-forming ISM distribution for metallicities of $Z=0.1$, $0.5$ and $2.0\,{\rm Z_{\odot}}$. Figures~\ref{fig:crt0p1}-\ref{fig:crt2p0} show the corresponding results for the star-forming ISM distribution.

\section{Using a different $A_{\rm V}/N_{\rm H}$ constant}
\label{app:avfac}

\begin{figure}
    \centering
    \includegraphics[width=0.49\textwidth]{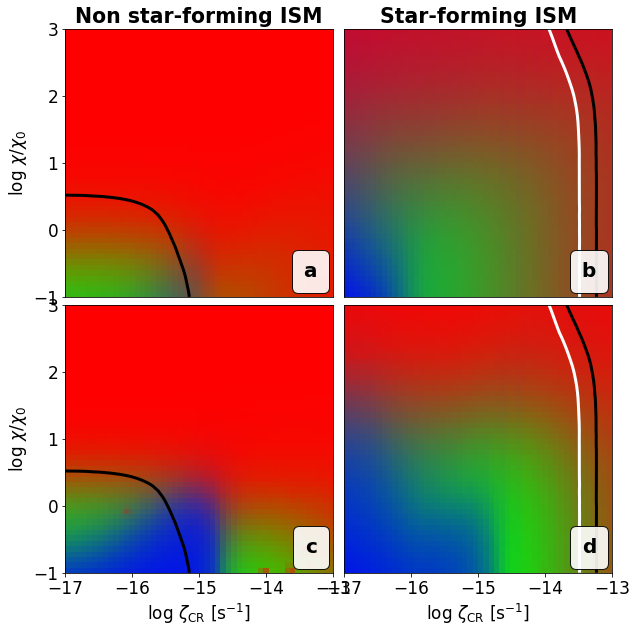}
    \caption{Carbon phases and H{\sc i}-to-H$_2$ transitions as discussed in Figs.~\ref{fig:lf} and \ref{fig:cf} for $A_{\rm V}/N_{\rm H}=5.3\times10^{-22}\,{\rm mag}\,{\rm cm}^2$. As can be seen, there are no appreciable differences in the results when adopting the lower $A_{\rm V}/N_{\rm H}$ that is frequently used in other similar PDR studies. In panels~(b) and (d), the white line corresponds to the LN-only distribution.}
    \label{fig:app_flags}
\end{figure}

As mentioned in the Introduction, in this work the conversion $A_{\rm V}=N_{\rm H}\cdot6.3\times10^{-22}\,{\rm mag}$ is used \citep{Roll07} for a more accurate comparison with \citetalias{Bisb19} and \citetalias{Bisb21}. However, the conversion $A_{\rm V}=N_{\rm H}\cdot5.3\times10^{-22}\,{\rm mag}$ \citep{Bohl78} is frequently used by other groups in PDR studies. Here, the grid of models for $Z=1\,{\rm Z}_{\odot}$ is repeated using the aforementioned constant, for both the non star-forming and the star-forming ISM distributions. Figure~\ref{fig:app_flags} illustrates the results. The $A_{\rm V}/N_{\rm H}=5.3\times10^{-22}\,{\rm mag}\,{\rm cm}^2$ constant lowers the range of $\chi/\chi_0-\zeta_{\rm CR}$ pairs for which the non star-forming distribution remains molecular, whereas it does not affect the H{\sc i}-to-H$_2$ transition in the star-forming distribution. The carbon cycle has no appreciable differences for either ISM distributions, both in terms of abundances and of antenna temperatures. It is, therefore, found that the slightly different $A_{\rm V}/N_{\rm H}$ ratio used does not alter the results in the present paper.

%%%%%%%%%%%%%%%%%%%%%%%%%%%%%%%%%%%%%%%%%%%%%%%%%%

% Don't change these lines
\bsp	% typesetting comment
\label{lastpage}
\end{document}